# Polaron-Driven Spin Funneling through Rashba-Split Bands in Mixed-Phase Quasi-Two-Dimensional Ruddlesden-Popper Perovskites

*Sushovan Sarkar [1,2], Koushik Gayen [1,2], Ashish Soni [1,2] and Suman Kalyan Pal [1,2]\**

[1]*School of Physical Sciences, Indian Institute of Technology Mandi, Kamand, Mandi-175005, Himachal Pradesh, India*

[2]*Advanced Materials Research Centre, Indian Institute of Technology Mandi, Kamand, Mandi-175005, Himachal Pradesh, India*

AUTHOR INFORMATION

**Corresponding Author**

\*E-mail: suman@iitmandi.ac.in; Phone: +91 1905 267040

## Abstract

Metal halide perovskites (MHPs) exhibit pronounced spin-orbit coupling (SOC) as a result of their heavy metal constituents, leading to distinctive electronic properties such as Rashba type band splitting which make them promising candidates for next generation spintronic applications. Here, using circularly polarized luminescence (CPL) and polarization dependent pump-probe spectroscopy, we found that spin polarization is present across all phases of our two-dimensional (2D) Ruddlesden-Popper (RP) mixed-phase perovskites, $(C_6H_7SNH_3)_2 (CH_3NH_3)_{n-1}Pb_nI_{3n+1}$ (n=1-4), irrespective of the number of inorganic layers. The origin of these spin polarized bands is attributed to the Rashba effect. Interestingly, the highly disordered nature of this system facilitates remarkably efficient ultrafast funneling of photoexcited spin-polarized excitons from the pure 2D phase (n=1) to higher-n phases at room temperature. We demonstrate that significant polaron formation due to the inherent soft crystal lattice and higher exciton-phonon interaction is responsible for the observed spin funneling effect in mixed-phase 2D RP perovskites. Polaron act as a protective mechanism for spin-polarized excitons, preserving their spin information through the screening of omnipresent phonon-induced momentum scattering. These findings not only offer valuable guidance for the design of 2D RP perovskites with pronounced Rashba effects but also unveil a compelling class of solution-processed perovskites capable of efficient spin-preserving energy transport at room temperature.

**Keywords:** Mixed-phase 2D-RP perovskite, Rashba splitting, Circularly polarized luminescence, Spin dynamics, Spin depolarization, Spin funneling, Pump-probe spectroscopy

## 1. Introduction

Semiconductor spintronics, which exploits the spin degree of freedom of electrons, holds promise for a new era of electronic devices. For spintronic applications to be effective, materials must enable efficient spin generation while also maintain a long spin polarization lifetime at room temperature to support reliable spin transport and detection.[1-4] However, achieving both efficient spin injection and long spin lifetimes simultaneously, remains difficult in conventional semiconductors. Typically, strong spin-orbit coupling (SOC) enhances spin injection efficiency but also accelerates spin relaxation, thereby shortening the spin lifetime.[5,6] As a result, reducing spin depolarization while maintaining high SOC remains a major challenge in advancing spintronic technologies. Pioneering studies on inorganic semiconductors such as GaAs, MnSe, and Si have encountered significant obstacles, including high costs and difficulties in device integration due to the need for high-vacuum and high-temperature processing, as well as strict lattice-matching constraints.[7-9] Organic semiconductors provide notable advantages, such as easy fabrication processes and structural tunability.[10, 11] Their intrinsically weak SOC is theoretically favorable for long-range spin current transport but also presents significant challenges for precise spin manipulation.[12] Furthermore, spin relaxation in these materials is primarily driven by hyperfine interactions, which severely restrict spin-relaxation lifetimes and limit their performance in spintronic applications.[11, 12]

Over the past decade, organic-inorganic metal halide perovskites (OIMHPs) with remarkable properties including tunable bandgap, high light absorption coefficient, high defect tolerance, long charge carrier diffusion length, and high photoluminescence quantum yield have shown great promise for photovoltaics, light-emitting diodes, lasers and photodetectors.[13-16] Following the rapid progress in optoelectronics, the potential of perovskites for spintronics applications has just begun to be investigated. It has been found that OIMHPs exhibit strong SOC due to the presence of heavy elements like lead (Pb) and halogens that determine the electron bands near their extrema points, which may lead to large Rashba splitting if the structure lacks inversion symmetry.[17, 18] Thus, the energy bands split into two branches with opposite spin orientations, producing spin-polarized charge carriers. Two-dimensional (2D) Ruddlesden-Popper (RP) perovskites, described by the general formula $(A')_2(A)_{n-1}Pb_nX_{3n+1}$, are structurally derived from three-dimensional (3D) perovskites by incorporating long-chain organic spacer cations (A′) between the inorganic $[PbX_6]^{4-}$ octahedral layers.[19] This results in a natural multiple quantum well (QW) configuration, where the parameter n indicates the number of stacked inorganic layers. This layered structure inherently breaks the spatial inversion symmetry present in the 3D counterpart.[20] As a result, structural distortions combined with strong SOC can produce Rashba splitted bands in the 2D RP perovskites and make them accessible for polarized spin generation and detection.[21] This characteristics make 2D perovskites highly appealing for spintronic applications.

Only a limited number of studies have so far investigated Rashba-type band splitting and spin depolarization mechanisms in 2D RP perovskite systems. In 2018, Chen et al. demonstrated that Rashba splitting reaches its maximum at n=4 in 2D RP perovskite single crystals, with a spin coherence lifetime of 7 ps at room temperature.[22] On the same time Mohammed and coworkers, through both theoretical and experimental work, revealed that Rashba splitting is present only in 2D RP perovskite single crystals with an even number of inorganic layers, and not in those with an odd number.[23] In stark contrast, Pham et al. found Rashba-Dresselhaus effect in 2D RP perovskite single crystals, reporting the maximum Rashba strength at n=1.[24] Additionally, Giovanni et al. observed an ultrafast spin funneling process in solution processed mixed-phase 2D RP perovskite from n=3 to n=4 phase and assigned this process to energy transfer through intermediated states, which preserves the spin and results in higher initial polarization.[25] However, the mechanism by which spin is preserved during energy transfer remains unclear. Recently, Yao et al. showed the circularly polarized electroluminescence (CPEL) is generated from a rapid energy transfer accompanied by spin transfer from 2D to 3D perovskites in chiral quasi-2D perovskite through chiral induced spin selectivity (CISS) effect.[26] While the energy or carrier transfer process in mixed-phase (i.e., phases having different layer numbers) 2D perovskites has been extensively investigated, the spin funneling phenomenon and its underlying mechanisms are not much explored in literature. Hence, the existence of the Rashba effect in 2D RP perovskites with specific n values is under debate, and the exact mechanism of spin funneling through the spin polarized bands in such mixed-phase perovskite system is still not well understood.

In this study, we systematically investigate spin polarization in 2D RP mixed-phase perovskites, $(TEA)_2(MA)_{n-1}Pb_nI_{3n+1}$ (n=1-4), exploiting both circularly polarized photoluminescence (CPPL) and polarization-resolved transient absorption (PRTA) spectroscopy. Remarkably, we observe a giant CPPL response with a polarization degree of 35% at 83 K from the n=1 phase, even without the application of an external magnetic field. Furthermore, our findings reveal the presence of spin polarized bands across all phases of the 2D RP mixed-phase perovskites, regardless of the number of inorganic layers (n). This result is further supported by circularly polarized TA measurements, which exhibit distinct signals under oppositely polarized pump-probe configurations. Our observations indicate that the inherent structural asymmetry arising from the coexistence of mixed phases in the samples enables Rashba splitting throughout all phases of the 2D RP perovskites. We propose that a spin funneling process is occurring in our mixed-phase sample, where spin-polarized carriers transfer from lower- to higher-n phases. This hypothesis is supported by measurements of the degree of spin polarization (DOSP), which was determined by selectively exciting higher-energy n phases in one of our mixed-phase samples. We suggest that the pronounced polaron formation, driven by the inherently soft crystal lattice and strong exciton-phonon interactions, is responsible for the observed spin-polarized exciton funneling in mixed-phase 2D RP perovskites. From our viewpoint, these findings provide a strategy for designing mixed-phase 2D RP perovskites that exhibit Rashba splitting and enable spin funneling in room temperature, thereby enhancing the performance of metal halide-based spintronic devices.

## 2. Experimental Section

### 2.1 Materials

The chemicals 2-thiopheneethylamine ($C_6H_7SNH_2$ (TEA), 98%), methyl ammonium iodide ($CH_3NH_3I$ (MAI), 99.9%), lead iodide ($PbI_2$, 99.999% purity, metals basis), dimethylformamide (DMF), hydroiodic acid (HI, 57%), dimethyl sulfoxide (DMSO), isopropyl alcohol (IPA), ethanol, diethyl ether and acetone were purchased from Sigma Aldrich. These chemicals were used directly without any additional purification.

### 2.2 Synthesis of 2-Thiopheneethylamine iodide (TEAI) salt

TEAI salt was synthesized by reacting TEA with HI, following the procedure described in our previous report.[27] Equimolar amounts of TEA and HI were dissolved in ethanol and stirred for 3 h. The resulting solution was then concentrated using rotary evaporation at 60°C under controlled conditions. The resultant TEAI salts were redissolved in an ethanol solution and repeatedly washed with diethyl ether to purify it. Finally, white crystals were obtained after drying at 60°C for 6 h and stored within the $N_2$ gas filled gloves box for further use.

### 2.3 Preparation of precursor solutions

The synthesis of the perovskite materials was performed using the methodology adopted in our earlier work.[28] For the synthesis of $(TEA)_2PbI_4$, 0.2 mmol of TEAI and 0.1 mmol of $PbI_2$ were dissolved whereas for $(TEA)_2MAPb_2I_7$, the precursor solution included 0.2 mmol of TEAI, 0.1 mmol of MAI, and 0.2 mmol of $PbI_2$. Similarly, for preparing $(TEA)_2MA_2Pb_3I_{10}$, the solution consisted of 0.2 mmol of TEAI, 0.2 mmol of MAI, and 0.3 mmol of $PbI_2$, while for $(TEA)_2MA_3Pb_4I_{13}$, the solution contained 0.2 mmol of TEAI, 0.3 mmol of MAI, and 0.4 mmol of PbI$_2$. All the salts were dissolved in a solvent mixture consisting of 900 μL of DMF and 100 μL of DMSO.

## 2.4 Preparation of perovskite films

Thin films of synthesized 2D perovskites were prepared via spin-coating as reported earlier.[28] Glass substrates were cleaned through a stepwise process involving ultrasonic treatment in detergent solution, deionized (DI) water, acetone and IPA for a duration of 10 min each. Subsequently, the substrates were subjected to UV-ozone treatment for 20 min. Precursor solutions of 2D perovskites were spin coated onto the treated glass substrates at a speed of 4000 rpm for 20 s, followed by annealing at 120°C for 5-6 min. We prepared four types of samples (S1-S4) to obtain 2D and quasi-2D perovskite phases: (TEA)$_2$PbI$_4$ in S1, (TEA)$_2$MAPb$_2$I$_7$ in S2, (TEA)$_2$MA$_2$Pb$_3$I$_{10}$ in S3, and (TEA)$_2$MA$_3$Pb$_4$I$_{13}$ in S4.

## 2.5 Characterization

Steady-state UV-visible absorption spectra were recorded using a Shimadzu UV-2450 spectrometer. The X-ray diffraction (XRD) patterns of (TEA)$_2$(MA)$_{n-1}$Pb$_n$I$_{3n+1}$ (n=1-4) perovskite films were obtained utilizing Rigaku Smart Lab 9 kW rotating anode diffractometer.

**Photoluminescence Spectroscopy**

In order to investigate CPPL phenomena, the mixed-phase 2D RP perovskite films were illuminated with 405 nm and 532 nm laser beams, which were sequentially passed through a polarizer and a quarter-wave plate. Photoluminescence (PL) from the sample was collected using a microscope objective (5X and 0.15 NA). For detection of CPPL, PL was first converted into linearly polarized light using a quarter-wave plate. An analyzer was then used to selectively transmit one polarization component at a time. To detect the other polarization component, the analyzer was rotated by 90°. Finally, PL intensity was recorded by spectrometer from Ocean optics (Model No.-USB4000) equipped with a charge-coupled device (CCD) detector, a grating (600 groves mm$^{-1}$) and an optical fiber.

**Helicity Resolved (HR) Femtosecond Transient Absorption (TA) Spectroscopy**

The TA spectroscopy system uses a femtosecond Ti: Sapphire amplifier (wavelength ~ 800 nm, output power 4 W, repetition rate 1 kHz, and pulse width < 35 fs).[29-31] The output from the amplifier (Spitfire Ace, Spectra physics) was then split into two beams, and the intense beam was sent to a nonlinear optical parametric amplifier (TOPAS-PRIME from Light Conversion) to generate a pump pulse. The probe beam was a white light continuum (WLC) generated by focusing the weaker part of 800 nm light (from Spitfire Ace) on a sapphire crystal for TA spectra measurements. The probe beam was detected in both with and without pump conditions with the help of a mechanical chopper of frequency 500 Hz. A stepper motor was controlled the time delay between pump, and probe pulses with an optical delay line. TA spectra were recorded by dispersing the beam with a grating spectrograph (Acton Spectra Pro SP 2358) followed by a CCD array. Light pulses of particular wavelengths from another TOPAS were used as probes while measuring TA kinetics. Two photodiodes having variable gain were used to record TA kinetics.[32, 33] We use Fresnel rhomb retarders (FR600QM), manufactured by Thorlabs, as quarter-wave plate for polarization dependent measurements. As the laser output from TOPAS is linearly polarized, we kept the optical axis of the retarder, FR600QM, at 45 ° for obtaining circular polarized light. During the experiment, we fixed the helicity of the pump beam and probe helicity was changed by using a Berek polarization compensator (variable wave plate, Model 5540, Newport) before the quarter wave plate.[34] Low-temperature measurements were carried out using a temperature-controlled sample holder (Linkam Scientific Instruments, Model No. LTS420E-PB4) cooled with liquid nitrogen.

## 3. Results and Discussion

**Circularly Polarized Luminescence**

The absorption and XRD spectra of the 2D perovskite samples were recorded at room temperature to confirm their structural and phase characteristics (**Figure S1**). The obtained results are consistent with our previous report,[28] which indicated that sample S1 is a pure 2D perovskite, whereas samples S2, S3, and S4 are quasi-2D perovskites comprising mixed phases. Among these, S2, S3, and S4 are predominantly composed of n=2, 3, and 4 phases, respectively. To investigate spin polarization in our 2D RP mixed-phase perovskites, CPPL measurements were conducted at 83 K and results are depicted in **Figure 1**. Distinct absorption peaks at 519 nm, 569 nm, 611 nm and 646 nm (**Figure S1**) are associated with the different layered phases from n=1 to 4, respectively. Therefore, in CPPL study, a 405 nm laser was selected for exciting the S1 sample, while for the excitation of S2, S3, and S4 samples, a 532 nm laser was employed. The observed PL peaks at 526 nm, 580 nm, 621 nm and 650 nm correspond to the phases from n=1 to 4, respectively. A noticeable intensity difference is observed between the right circular ($\sigma^+$) and left circular ($\sigma^-$) PL across all phases of S1 and S2 samples (**Figure 1**a, b). In contrast, as shown in **Figure 1**c, the S3 sample exhibits minimal difference between $\sigma^+$ and $\sigma^-$ PL emission, with identical PL intensity for both polarizations in the n=4 phase. On the other hand, in S4 sample, there are no differences between $\sigma^+$ and $\sigma^-$ PL emission of n=2-4 phases, except in the bulk phase (775 nm), where a slight disparity is still evident (**Figure 1**d). To quantify the degree of the CPPL, we introduce a parameter, defined as[35, 36]

$$P_{CPPL} = \frac{I(\sigma^+) - I(\sigma^-)}{I(\sigma^+) + I(\sigma^-)} \quad \text{...................} \quad (1)$$

where I ($\sigma^+$) and I ($\sigma^-$) denote the intensity of the $\sigma^+$ and $\sigma^-$ PL, respectively. We determined $P_{CPPL}$ focusing only on the main PL peak of each sample (i.e., at 526 nm for n=1 in S1, 580 nm for n=2 in S2, 621 nm for n=3 in S3 and 650 nm for n=4 in S4). Remarkably, the maximum $P_{CPPL}$ value for the n=1 phase of S1 reaches 35% at 83 K, approximately twice the reported value from the chiral (S-MBA)$_2$PbI$_4$ at 77 K and 7 times greater than that in the (BA)$_2$PbI$_4$ at 10 K, which was 4.8% under zero magnetic field.[24, 37] Furthermore, in the n=2 phase of the S2 sample, the value of $P_{CPPL}$ is ~30%, whereas in the n=3 phase of the sample S3, it is observed to be ~6% at 83 K. This reduced value of $P_{CPPL}$ in the n=3 phase is attributed to thermal relaxation effects following non-resonant excitation, which tends to distribute carriers more evenly between the two spin polarized states compared to excitation near resonance.[38] It is surprising that the degree of polarization in the n=2 and 3 phases remains remarkably high at 83 K, which stands in stark contrast to earlier report on 2D RP single crystal, $(BA)_2(MA)_{n-1}Pb_nI_{3n+1}$, where the polarization for the same phases were found to be nearly absent even at 10 K.[24] Nonetheless, Yin et al. reported that the Rashba splitting is observed in $(PEA)_2(MA)_{n-1}Pb_nI_{3n+1}$ single crystals only when the number of inorganic layers is even, with no such splitting detected for odd-layered structures.[23] This observation, however, does not align with our findings. We propose that this discrepancy arises from the presence of mixed phases in our samples, which inherently introduce structural asymmetry irrespective of the number of inorganic layers, thereby enabling Rashba splitting across all phases of 2D RP perovskites.

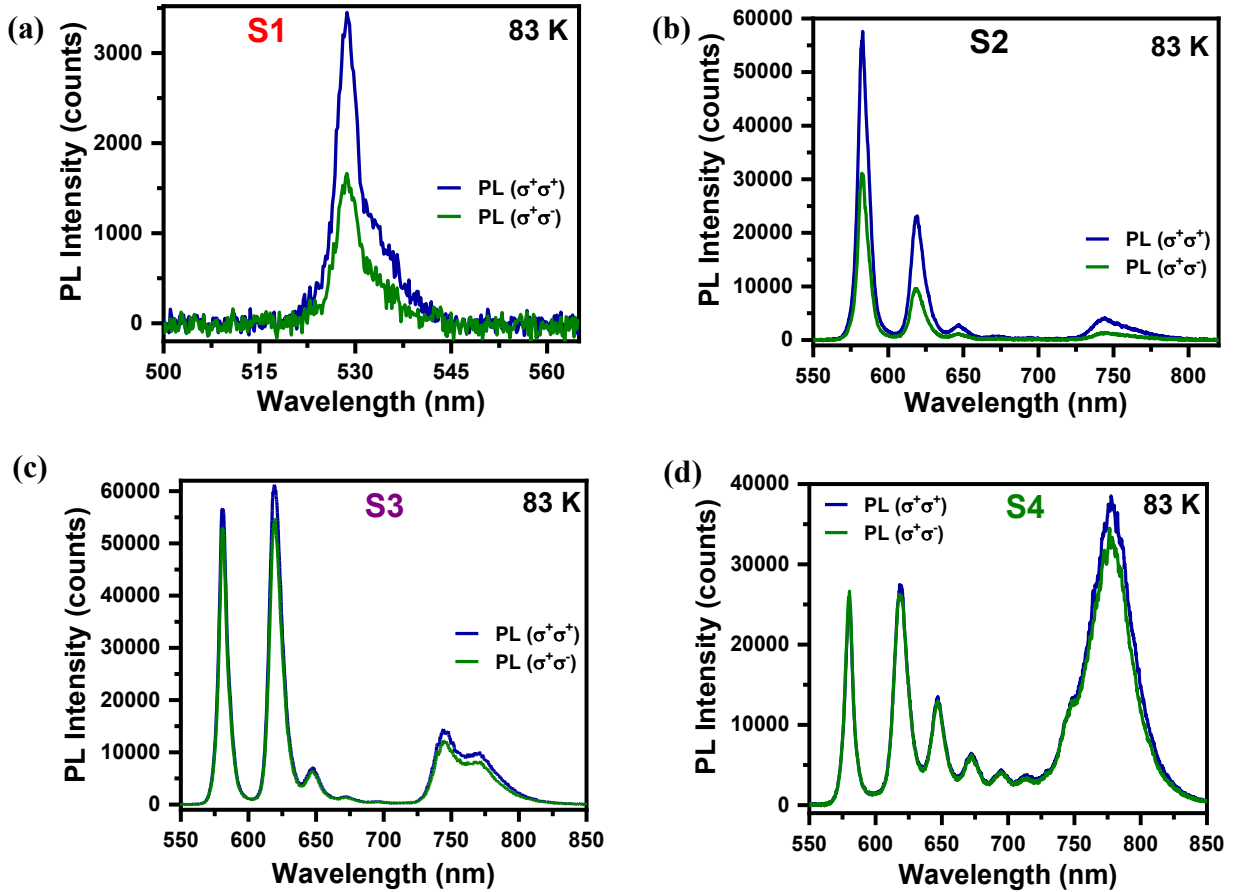

**Figure 1(a-d)** Polarization-sensitive PL spectra of (TEA)$_2$(MA)$_{n-1}$Pb$_n$I$_{3n+1}$ (n=1-4) thin films at 83 K. Panels a-d correspond to samples S1-4, respectively. Right circularly ($\sigma^+$) polarized laser lights of wavelengths 405 nm (for S1 sample) and 532 nm (for S2-4) were used for excitation.

**Figure S2**(a-c) shows the temperature dependence of CPPL response, defined as $\Delta I = [I(\sigma^+) - I(\sigma^-)]$, for the main phases of samples S1-S3. Additionally, temperature-dependent behavior of $P_{CPPL}$ was studied over a temperature range of 83 to 298 K [**Figure S2**(d-f)]. The $P_{CPPL}$ value of the n=1 phase remained nearly constant at approximately 30% in the temperature range of 100-220 K and only started to decrease sharply near 240 K, at which the structural phase transition occurs.[39] In the n=2 phase, the $P_{CPPL}$ value increases around 40% at 93 K and almost remains same up to 160 K. In contrast, the n=3 phase exhibits a fluctuating $P_{CPPL}$ behavior, varying between 3-4% over the same temperature range. It should be noted that the significant CPPL response is observed in the n=1 phase up to 260 K whereas for n=2 and 3, it persists till 170 K only. Beyond these temperatures, though the PL intensity remains relatively

high, the CPPL response is roughly zero, consistent with earlier reports, linking such suppression to structural phase transformation.[24, 40] With temperature variation, the structural phase of the perovskite evolves, which may lead to the restoration of inversion symmetry in the crystal lattice. Under such conditions, Rashba splitting is no longer present, leading to the disappearance of $P_{CPPL}$. This suggests that the CPPL mechanism in 2D RP perovskites is highly sensitive to the crystal structure of the material and its symmetry properties.

**Spin Polarization Dynamics**

We further used circular polarization-resolved TA spectroscopy to probe spin-polarized bands and resulting spin polarization in the 2D RP mixed-phase perovskites, $(TEA)_2(MA)_{n-1}Pb_nI_{3n+1}$ (n=1-4). **Figure 2** (a-d) shows TA spectra of the samples S1-S4, respectively recorded at a delay time of 0.5 ps under circularly polarized pump and probe lights. In all the measurements, the pump light was polarized in the $\sigma^+$ direction, while the broadband probe light was configured as $\sigma^+$ or $\sigma^-$ polarization to achieve the same (SCP) or opposite (OCP) circularly polarized TA detection, respectively. In both SCP and OCP configurations, the spectrum of the S1 sample exhibits a ground state bleach (GSB) signal at 520 nm, which arises due to state filling effect and corresponds to the excitonic absorption of the n=1 phase (**Figure 2**(a)). For the S2 and S3 samples, additional GSB signals appear at 576 nm and 620 nm, associated with the excitonic absorption of the n=2 and n=3 phases, respectively (**Figure 2**(b-c)). In the case of the S4 sample, TA spectra include more GSB bands corresponding to the absorption of the n=4 phase at 660 nm, as well as for the bulk phase centered around 750 nm (**Figure 2**(d)). A stronger SCP signal relative to the OCP signal suggests an imbalance of carrier populations in different spin states, originating from spin-selective excitation. When the probe polarization matches (or is counter to) the pump polarization, the state filling signal corresponds to the majority (or minority) spin population. The convergence of these two state-filling signals over time reveals spin polarization relaxation dynamics. All these findings again lead us to conclude that spin polarized bands are present in all the phases of our 2D RP mixed-phase perovskites.

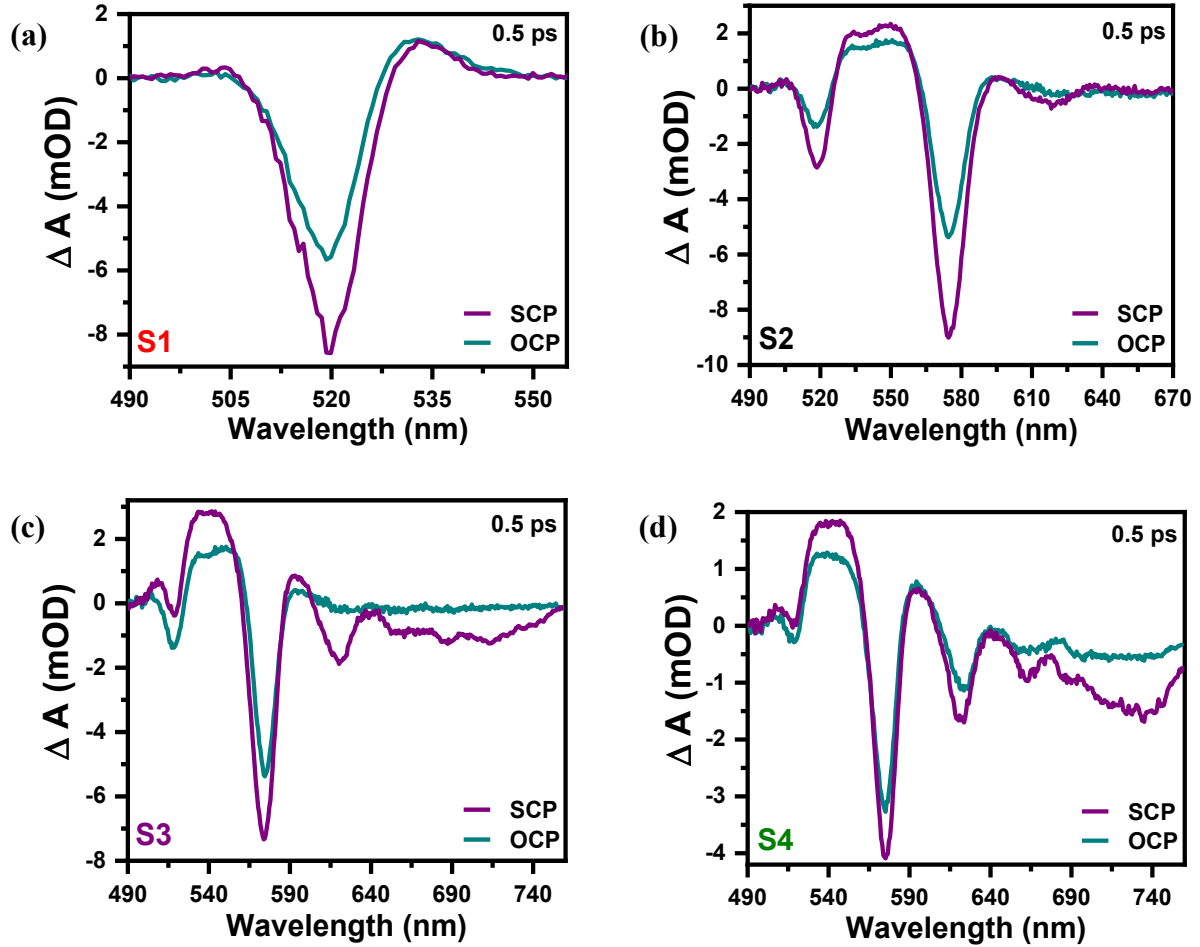

**Figure 2 (a-d)** The SCP and OCP TA spectra for S1-S4 samples, respectively measured at room temperature at a pump-probe delay of 0.5 ps. The pump wavelength was 480 nm.

To investigate the spin polarization dynamics in our samples, we measure the exciton bleach kinetics under band-edge excitation corresponding to the main phases of all the samples using HRTA spectroscopy. **Figure 3** (a-d) illustrates the TA kinetics of n=1-4 phases, respectively. It is evident from the figures that the spin flip induces a concurrent decay of the bleach signal corresponding to σ⁺ probe and the formation of the signal for σ⁻ probe. The decay and formation kinetics converge at a later time, signifying the depolarization of the initially polarized spin population. The DOSP can be defined as[41]

$$P = \frac{\Delta A_{scp} - \Delta A_{ocp}}{\Delta A_{scp} + \Delta A_{ocp}} \quad \ldots\ldots\ldots\ldots\ldots\ldots\ldots (2)$$

where $\Delta A_{scp}$ and $\Delta A_{ocp}$ represent the TA signal in SCP and OCP modes, respectively. We have calculated DOSP from polarization dependent TA kinetics of all the phases **Figure 3** (a-d).

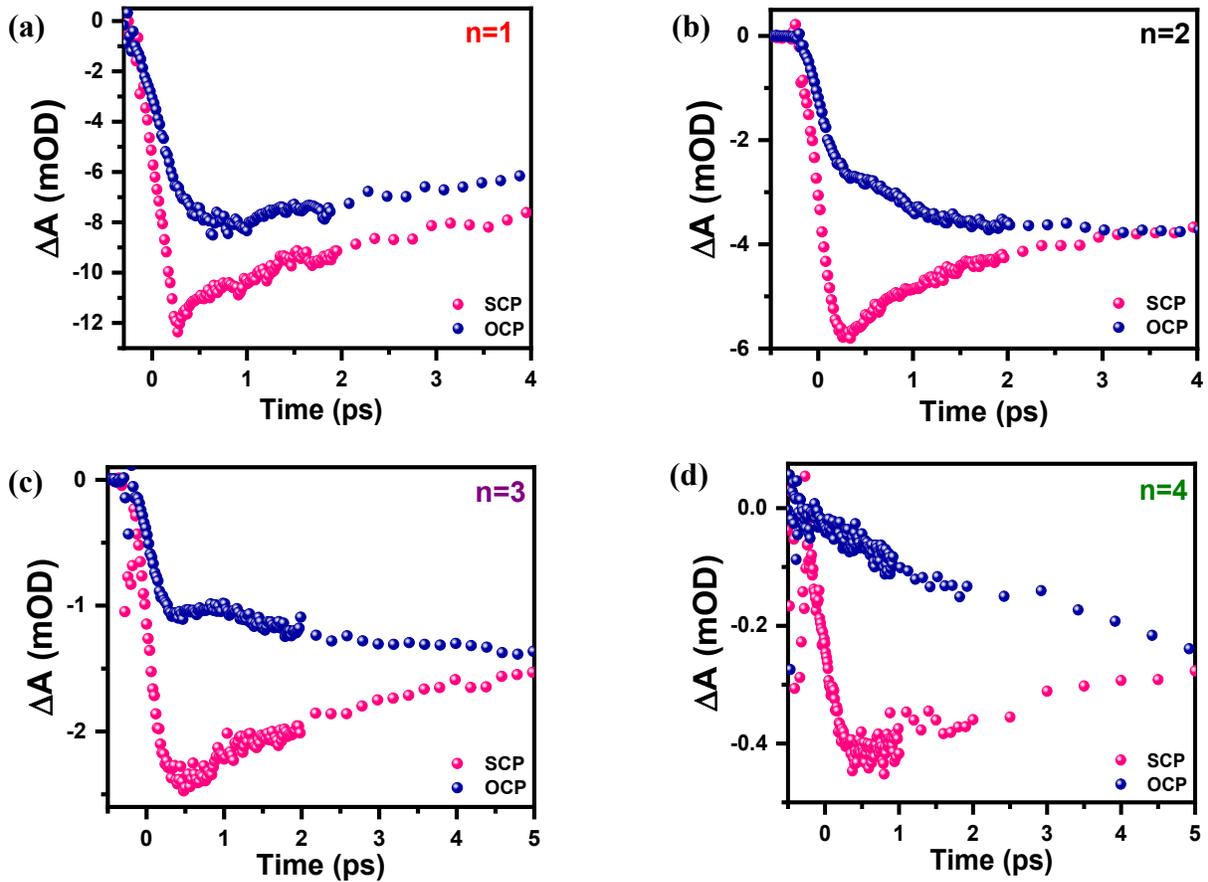

**Figure 3** TA kinetics measured under band-edge excitation with same (SCP)- and opposite (OCP)-circularly polarized pump-probe configurations at the exciton bleach positions of **(a)** n=1 (520 nm), **(b)** n=2 (570 nm), **(c)** n=3 (620 nm), and **(d)** n=4 (670 nm) phases. The excitation

wavelengths were 510 nm for n=1, 560 nm for n=2, 595 nm for n=3 and 640 nm for n=4 phases. Pump fluences were 8.7-12.6 µJ/cm$^2$.

We found that at room temperature, the maximum DOSP for the n=1 and n=2 phases are approximately 25% and 35%, respectively. In stark contrast to the CPPL results, the maximum DOSP values for n=3 and n=4 phases reach 36% and 71% respectively under band-edge excitation at room temperature. This discrepancy can be attributed to the fact that at far-band-edge excitation, carriers undergo thermal relaxation, which reduces spin polarization due to momentum scattering processes.

**Spin Funneling**

We further propose that a spin funneling phenomenon may occur, where spin-polarized carriers in Rashba-split states are transferred from lower-n to higher-n phases. This hypothesis is inspired by previous reports that have demonstrated energy and carrier transfer processes in mixed-phase RP perovskite systems.[42-44] Building on these findings, we speculate that spin funneling could similarly take place in such mixed-phase RP perovskite systems. To validate our hypothesis, we measured the circular polarization dependent TA kinetics of the S3 sample at the GSB position of n=3 phase (620 nm) using a $\sigma^+$ pump following band-edge excitation of n=1, 2 and n=3 phases (**Figure 4**(a-c)). As previously mentioned, under band-edge excitation (~595 nm) the n=3 phase exhibits a maximum DOSP value of 36%. However, when the S3 sample is excited with a 560 nm pump pulse, the initial spin polarization increases to 38.8%. With a further reduction of excitation wavelength to 510 nm, the maximum DOSP value reaches 47.4%. We attribute the observed progressive enhancement in DOSP with enhancing excitation energy to the efficient funneling of spin-polarized excitons from the Rashba split state of the lower- n phases, which were selectively populated. The spin-polarized excitons created in the n=1 phase are transferred via the intermediate n=2 phase and ultimately accumulate in the corresponding spin state of the lower-energy n=3 phase. This sequential spin-conserving transfer enhances the population imbalance between the two oppositely spin-polarized Rashba-split states in the n=3 phase, thereby resulting in an enhanced DOSP.

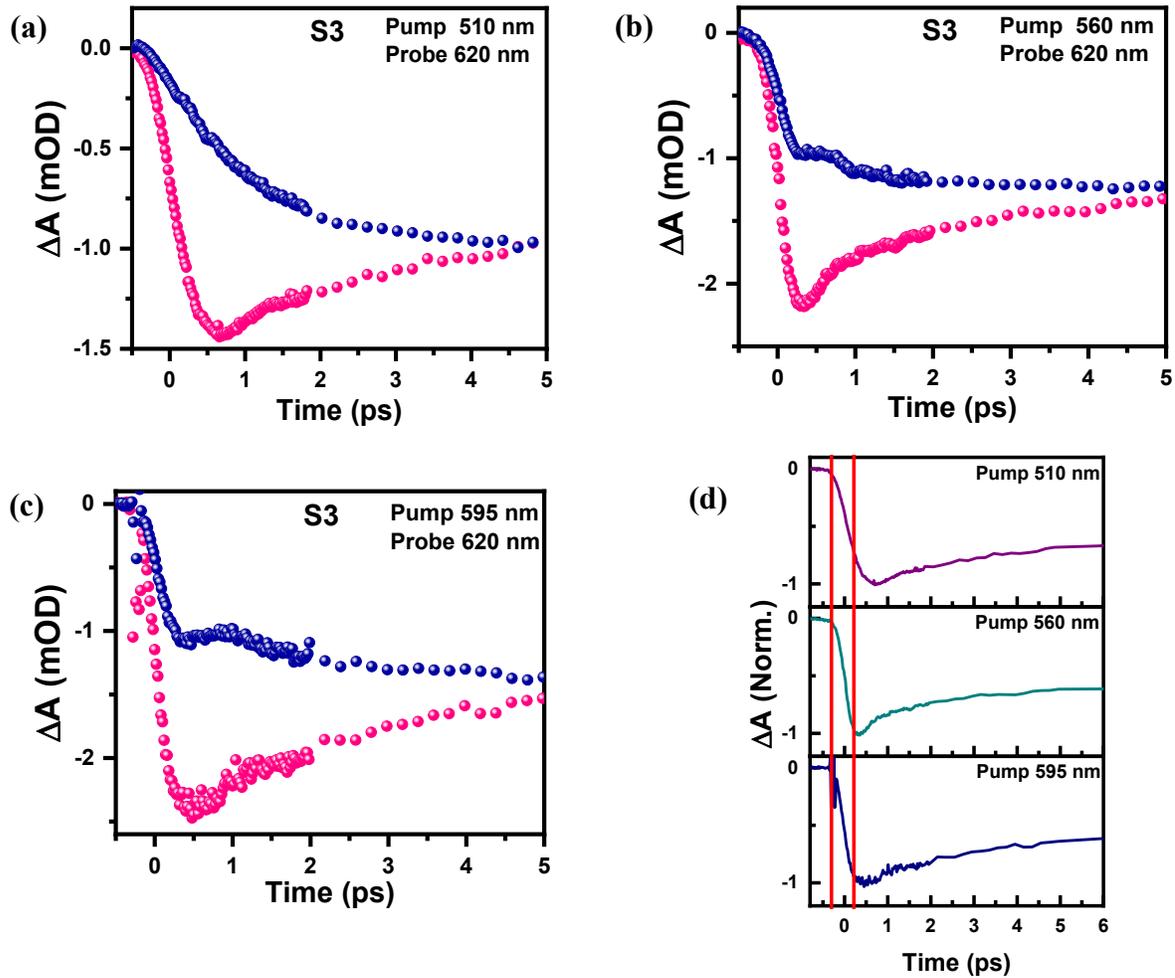

**Figure 4 (a-d)** TA kinetics of the S3 sample measured at the bleach position of the n=3 phase under band-edge excitation of (a) n=1, (b) n=2, and (c) n=3 phases. The pink and blue symbols correspond to SCP and OCP, respectively. (d) Rise dynamics of the n=3 bleach signal under various excitation conditions, showing a clear increase in rise time with higher pump photon energy. Red vertical lines serve as visual guides.

Additionally, we performed a similar study for the bulk phase having TA bleach at 760 nm (**Figure S3**(a-d)). In contrast to the behavior observed in the n=3 phase, the initial spin polarization in the bulk phase exhibits an opposite trend when excited at different higher energy n phases. Under band-edge excitation of the bulk phase at 730 nm, the maximum DOSP is found to be 32%. However, this value drops sharply to 10.3% when excitation is tuned to the band-edge of the n=4 phase (630 nm). A further gradual decline in DOSP is observed with exciting lower-n phases, reaching 5.1% and 3.8% for the n=3 (590 nm) and n=2 (560 nm) phases, respectively. Ultimately, no measurable spin polarization is detected under band-edge excitation of the n=1 phase (510 nm). In conventional semiconductors, under nonresonant photoexcitation, hot carriers or excitons undergo thermalization toward the band-edge states through momentum-scattering process, as is also evident in pure 3D perovskites and the bulk phase of our sample.[4, 7] These momentum-scattering events inherently contribute to loss of spin information, thereby causing a reduction in the DOSP. Therefore, the different trend in the bulk phase suggest that the enhanced spin polarization observed in the n=3 phase of the S3 sample primarily arises from the sequential transfer of spin-polarized excitons from lower-n phases, rather than from the thermalization of photoexcited hot carriers or excitons within the n=3 phase itself. Also, we confirm our result by comparing the signal rise time upon different pump excitation. Specifically, we observe an increase in the rise time of n=3 phase from 0.2 ps at 595 nm excitation to 0.7 ps at 515 nm excitation as shown in **Figure** 4d. In contrast, the bulk phase exhibits a much faster and nearly excitation-independent thermalization time of approximately 0.35 ps. This discrepancy indicates that the enhanced spin polarization in n=3 phase for higher energy excitations is not predominantly governed by the thermalization of photoexcited hot carriers.

**Mechanisms of Spin Depolarization and Spin Information Retention**

To investigate how spin-information is retained during the exciton funneling from lower to higher-n phase and to uncover the underlying mechanism behind spin depolarization in different phases we conducted circular polarization-resolved TA kinetics measurements under band-edge excitation, systematically varying both temperature and pump fluence. The difference between the SCP and OCP TA kinetics ($|\Delta A_{scp}-\Delta A_{ocp}|$) governs the spin relaxation dynamics in the 2D perovskites (**Figure 5**). Spin lifetime of each phases was calculated by fitting the corresponding decay kinetics by a single-exponential function (**Figure 5**) and presented in **Table-S1**. The n=1 phase reveals an ultrafast spin lifetime of approximately 0.35 ps at room temperature (**Table-S1**), in agreement with previous reports on n=1 2D MHPs.[22, 41, 45] Such rapid spin depolarization is expected, given the exceptionally large exciton binding energy ($E_b$~237 meV) in n=1 2D MHPs,[28] which leads to a strong Coulomb exchange interaction. This behavior is consistent with the Maialle-Silva-Sham (MSS) mechanism of spin relaxation. Furthermore, due to the weak temperature dependence of the Coulomb interaction,[46] the spin relaxation time exhibits negligible variation across the entire temperature range (83-298 K), as shown in **Table-S1**. At room temperature, the n=2 and n=3 phases exhibit nearly identical spin lifetimes of 2.04 ps and 1.83 ps, respectively, while the n=4 phase displays a slightly longer spin lifetime of 2.4 ps (**Table-S1**). These values are significantly longer than that observed for the n=1 phase, underscoring the influence of quantum well thickness on spin relaxation dynamics. As the $E_b$ decreases with increasing layer number, the Coulomb exchange interaction becomes progressively weaker in the n=2 to n=4 phases compared to the n=1 phase. Consequently, slower spin decay is expected in phases with higher n values. Moreover, the spin relaxation time of the n=2-4 phases exhibit a noticeable temperature dependence, decreasing with lower temperatures,

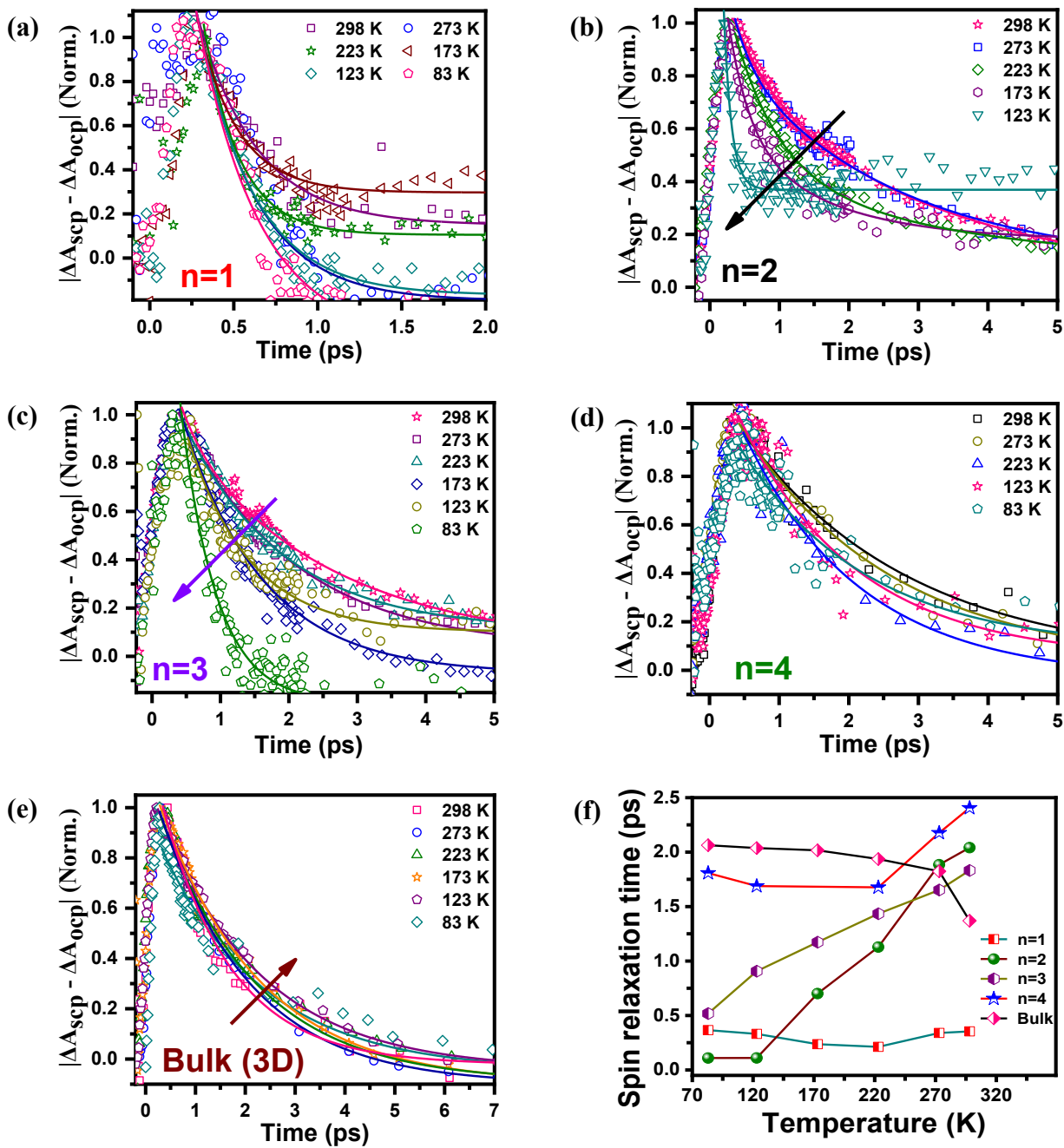

**Figure 5** Spin relaxation kinetics of **(a)** n=1, **(b)** n=2, **(c)** n=3 **(d)** n=4, and **(e)** bulk (3D) phases of the quasi-2D perovskite measured over a temperature range of 83-298 K. Solid lines indicate

single-exponential decay fits to the data. **(f)** Variation of spin lifetime with temperature at different phases of the quasi-2D perovskite.

as shown in **Table-S1**. This trend suggests the involvement of an additional spin depolarization mechanism beyond the Coulomb exchange interaction. To gain insight into the underlying mechanism, spin lifetimes of different phases are plotted as a function of temperature in **Figure 5f**. For the n=2 phase, the spin lifetime sharply drops from 1.88 ps to 0.11 ps as the temperature decreases from 273 K to 123 K, followed by a plateau at 0.1 ps between 123 K and 83 K. In the n=3 phase, the spin lifetime gradually reduces from 1.83 ps to 0.91 ps as the temperature lowers from 298 K to 123 K, followed by a sudden drop to 0.51 ps upon further cooling to 83 K. In contrast, the n=4 phase exhibits a different trend, with the spin lifetime decreasing from 2.4 ps to 1.7 ps at 223 K and then remaining nearly constant down to 83 K. In conventional semiconductors, the dominant spin relaxation mechanisms typically include the D'yakonov-Perel' (DP), Elliott-Yafet (EY), Bir-Aronov-Pikus (BAP), and MSS mechanisms. The DP mechanism typically dominates in systems lacking inversion symmetry, where spin depolarization arises from SOC via an effective internal magnetic field. As SOC is largely insensitive to temperature, the DP mechanism generally exhibits weak temperature dependence.[47] In contrast, the EY mechanism is associated with spin relaxation via momentum scattering of carriers, which becomes less effective at lower temperatures, thereby extending the spin lifetime.[47, 48] Based on our observations that the spin relaxation time in the n=2 and n=3 phases decrease upon lowering temperature, we can probably rule out the DP and EY mechanism for these phases. As noted earlier, the MSS mechanism, driven by Coulomb exchange interaction, is also expected to exhibit weak temperature dependence.[46] Therefore, the strong temperature sensitivity of the spin depolarization observed in the n=2 and n=3 phases suggest the involvement of an additional spin relaxation mechanism. Recent studies on spin dynamics in perovskite materials have demonstrated that, in addition to conventional spin-relaxation mechanisms, an alternative relaxation pathway mediated by polaronic states also plays a significant role.[41, 49] In contrast to traditional inorganic semiconductors, MHPs possess a soft and polar crystal lattice with large anharmonicity, which results in strong exciton-phonon coupling.[50-52] Consequently, charge carriers in perovskites become dressed by local lattice distortions, forming large polarons that shield them from impurity scattering and electron-hole recombination.[52-54] This polaronic protection facilitates an alternative spin-relaxation pathway, which can dominate the spin dynamics of exciton-polarons in MHPs. Specifically, at higher temperatures, the formation of polaronic states is more favorable, resulting in reduced overlap of exciton wave functions and thus a longer spin lifetime. As the temperature decreases, phonon activity diminishes, weakening the exciton-phonon coupling and the polaronic states cannot be efficiently formed. However, at low-temperatures, the MSS mechanism becomes the dominant spin relaxation pathway, resulting in a shortened spin lifetime. This behavior aligns well with the observed temperature dependence of the spin lifetime in the n=2 and n=3 phases (**Figure 5**f). Moreover, as the temperature decreases, the bulk phase exhibits a pronounced increase in spin lifetime, which shows an opposite trend from those observed in the n=2, 3, and 4 phases (**Figure 5**f). This behavior suggests that the impurity scattering driven EY mechanism predominantly governs spin relaxation in the bulk phase.

To further understand the role of polaronic states in spin relaxation and retention of spin information during spin funneling, we carried out pump fluence-dependent TA measurements on the n=1-4 phases at room temperature over a pump fluence range of approximately 0.9 to 34.7 µJ/cm² (**Figure 6**). Spin lifetimes were estimated by fitting the spin dynamics of each cases by a single-exponential decay function (**Table-S2**). **Figure 6**a presents the spin decay curves for the n=1 phase, showing a saturation trend with increasing pump fluence. The ultrafast spin lifetime, which exhibits minimal dependence on pump fluence, can still be attributed to the MSS mechanism, consistent with the exceptionally high $E_b$ in the n=1 phase. As shown in **Figure 6**b, the spin lifetime of the n=2 sample decreases rapidly and consistently with increasing pump fluence. In contrast, the n=3 phase maintains a more gradual and slow reduction in spin decay lifetime as the pump fluence increases (**Figure 6**c and **Table-S2**). For the n=2 phase, the spin lifetime shows a pronounced sensitivity to pump fluence, decreasing from 2.8 ps at 0.9 µJ/cm² to 0.7 ps at 15.9 µJ/cm² (**Figure 6**f). Though this behavior aligns with the MSS mechanism, unlike in conventional III-V quantum wells and transition metal dichalcogenides, the dependence of spin lifetime on pump fluence here is nonlinear.[41, 55] An unusually long spin lifetime of 2.8 ps is observed at the lowest pump fluence. This anomaly suggests that another mechanism, namely the formation of polaronic states, plays a significant role in the spin relaxation dynamics of the n=2 phase at low fluences. At low exciton densities, the reduced exciton density limits the overlap of exciton wave functions, thereby weakening the Coulomb exchange interaction. This, combined with the enhanced formation of polaronic states, results in a markedly prolonged spin lifetime. In contrast, at higher fluences, the Coulomb exchange interaction becomes the dominant factor, diminishing the influence of polaronic effects. As a result, the MSS mechanism dominates, leading to a significantly shorter spin lifetime. For the n=3 phase, the spin lifetime shows a slightly different picture, gradually decreasing from 2.4 ps to 1.4 ps as the pump fluence increases from 5.3 µJ/cm² to 31.9 µJ/cm² (**Figure 6**f). Unlike the n=2 phase, $E_b$ in the n=3 phase is further reduced to around 188 meV, leading to a weaker Coulomb exchange interaction.

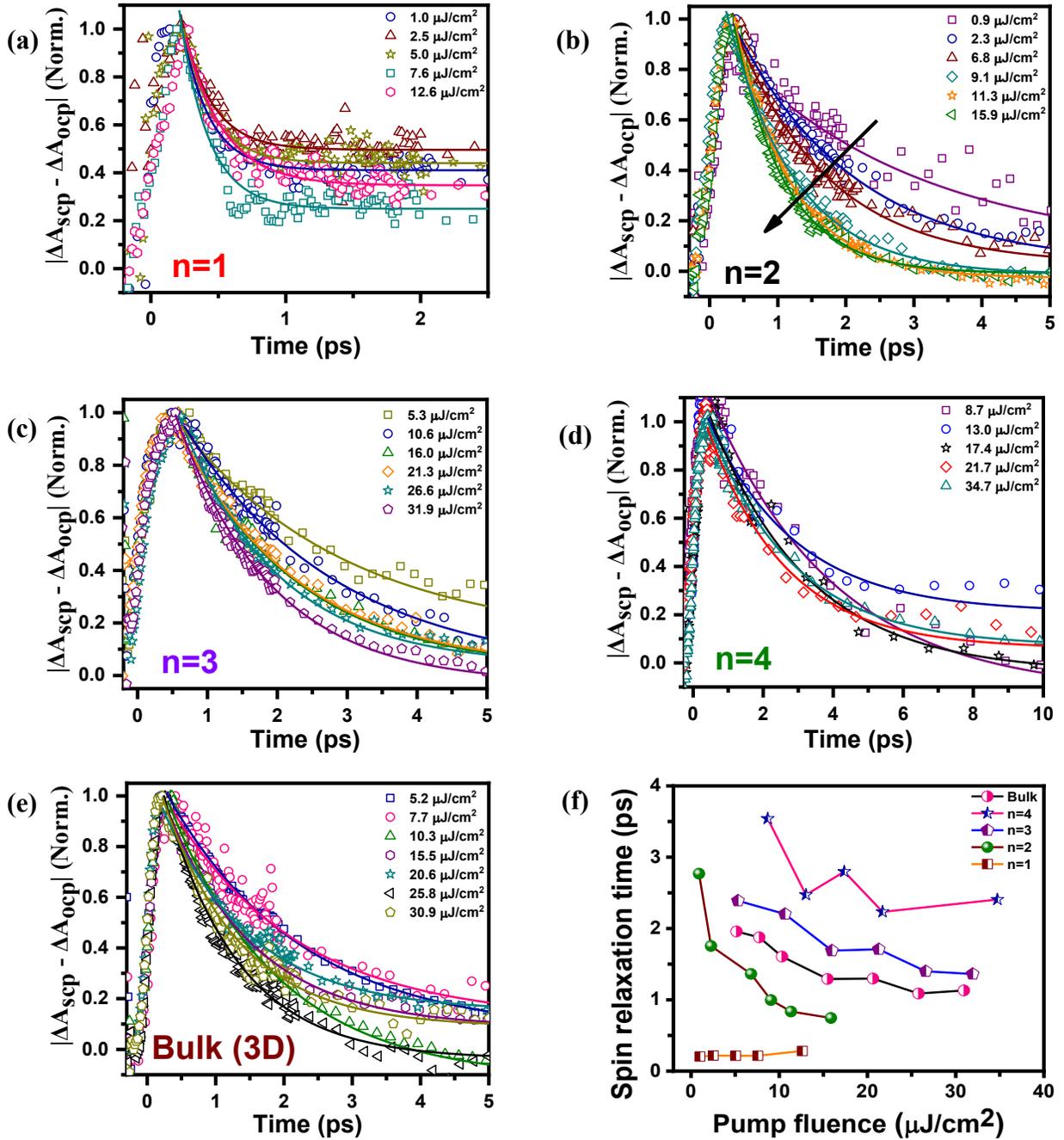

**Figure 6** Spin relaxation kinetics for **(a)** n=1, **(b)** n=2, **(c)** n=3, **(d)** n=4, and **(e)** bulk (3D) phases of the quasi-2D perovskite under different pump fluences at room temperature. The solid curves represent exponential decay that fits the data. **(f)** Variation of spin lifetime with pump fluence in different phases of the quasi-2D perovskite.

Consequently, polaronic effects are likely to play a more prominent role in governing spin relaxation in this phase. In contrast, the n=4 phase shows a completely different trend, where the spin lifetime abruptly dropping from 3.5 ps at 8.7 μJ/cm$^2$ to 2.5 ps at 13.0 μJ/cm$^2$ and then remain nearly constant at around 2.4 ps over the higher fluence range (17.4 to 34.7 μJ/cm$^2$). As the $E_b$ of n=4 phase is further reduced to 60 meV, the Coulomb exchange interaction significantly screened even at high pump fluences. Moreover, our previous study revealed that as the number of inorganic layers (n) increases, the exciton-phonon interaction becomes stronger, thereby amplifying the polaronic effect.[28] As a result, polaron-mediated processes emerge as the predominant mechanism governing spin depolarization in these systems. Lastly, in the bulk phase, the low value of $E_b$ facilitates the formation of free charge carriers, which subsequently form polarons. These free carrier-polarons are more prone to scattering by charged impurities and grain boundaries.[56] As shown in **Figure 6**e, the observed monotonic decrease in spin relaxation time with increasing pump fluence provides additional evidence that the spin relaxation is governed by the EY mechanism.

## 4. Conclusions

In summary, we have experimentally investigated spin polarization in our mixed-phase 2D-RP perovskites through CPPL, and circular polarization resolved TA spectroscopy. The remarkably strong CPPL responses-especially for n=1 and 2 phases, with $P_{CPPL}$ values of approximately 35% and 30% respectively-along with the distinct differences observed in TA signals under two opposite circular pump-probe configurations, collectively indicate the presence of Rashba-split bands in all phases of the 2D RP perovskites, irrespective of the number of inorganic layers (n). Due to the coexistence of multiple phases in our samples, structural asymmetry is inherently present in the system, which facilitates Rashba splitting across all phases of the 2D RP perovskites. Additionally, we have demonstrated the occurrence of a spin funneling process in our mixed-phase sample, wherein spin-polarized carriers transfer from lower- to higher-n phases. This interphase transfer leads to an increase in the DOSP value of n=3 phase from 36% to 47.4% when excited at the band edges of the n=3 and n=1 phases, respectively. We revealed that the inherent soft crystal lattice combined with strong exciton-phonon interaction leads to substantial polaron formation in the system. As the $E_b$ decreases and exciton-phonon interaction strengthens with increasing inorganic layer number, then the polaronic effect becomes progressively more prominent from n=1 to n=3 phase. These polarons act as a protective shield for the photogenerated carriers in the higher-energy n phases, helping to preserve their initial spin-polarized state during the funneling process from n=1, 2 to the n=3 phase. Moreover, in the bulk phase, spin relaxation predominantly occurs through momentum scattering processes, which lead to the loss of spin information and consequently result in a reduced DOSP value when the system is excited at higher-energy n phases. These findings reveal the existence of Rashba-split bands in mixed-phase 2D-RP perovskites, establishing them as an effective platform for room-temperature spin funneling and highlighting their promising potential for the advancement of next-generation perovskite spintronic devices.

## Acknowledgements

The authors gratefully acknowledge the Advanced Materials Research Centre (AMRC) at the Indian Institute of Technology Mandi for providing the necessary experimental facilities. S.S. and K.G. also thank the Ministry of Education (MOE) for fellowship support.

# References

# Supporting Information

**Polaron-Driven Spin Funneling Through Rashba-Split Bands in Mixed-Phase Quasi-Two-Dimensional Ruddlesden-Popper Perovskites**

*Sushovan Sarkar [1,2], Koushik Gayen [1,2], Ashish Soni [1,2], and Suman Kalyan Pal [1,2]*

[1]School of Physical Sciences, Indian Institute of Technology Mandi, Kamand, Mandi-175005, Himachal Pradesh, India

[2]Advanced Materials Research Centre, Indian Institute of Technology Mandi, Kamand, Mandi-175005, Himachal Pradesh, India

AUTHOR INFORMATION

**Corresponding Author**

*E-mail: suman@iitmandi.ac.in; Phone: +91 1905 267040

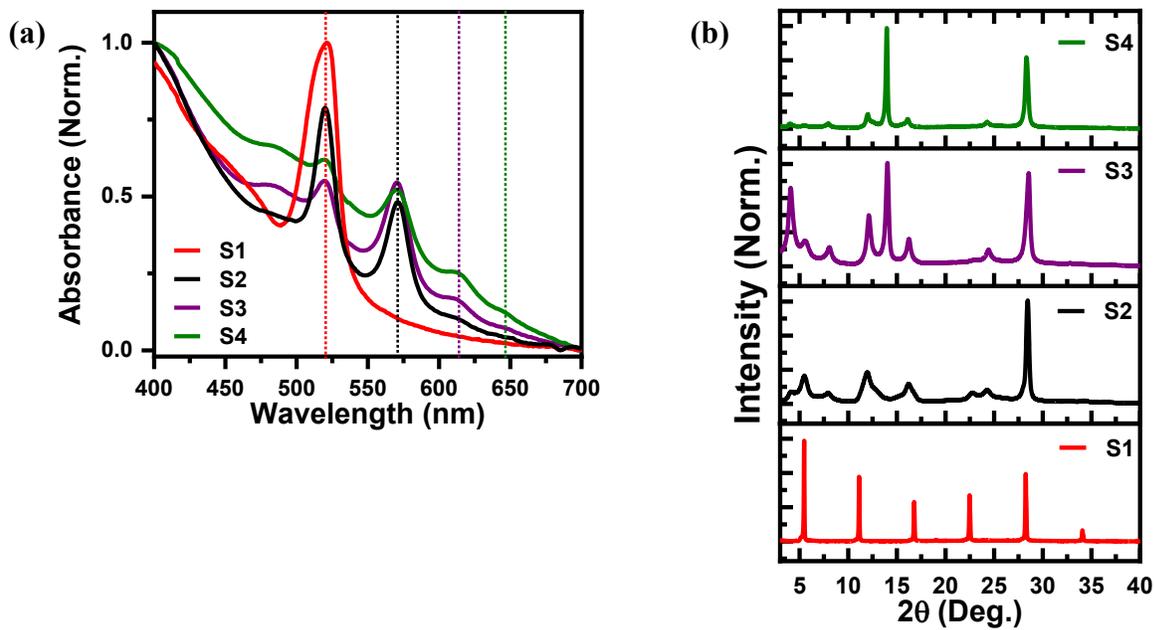

**Figure S1 (a)** Steady state absorption spectra of 2D-RP perovskites films S1-4. **(b)** X-ray diffraction (XRD) patterns of $(TEA)_2(MA)_{n-1}Pb_nI_{3n+1}$ (n=1-4) perovskite films.

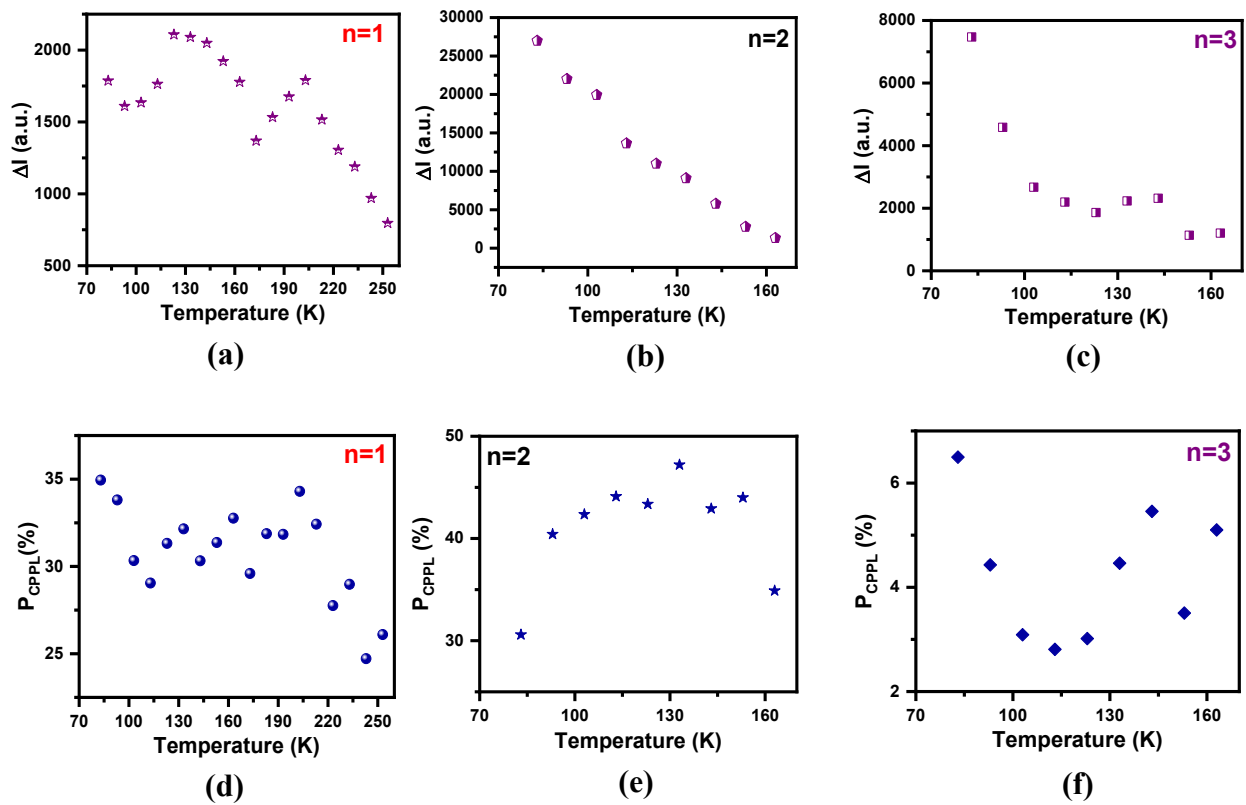

**Figure S2** Temperature-dependent **(a-c)** CPPL response where ΔI = [I (σ⁺) - I (σ⁻)] and **(d-f)** degree of the CPPL, P$_{CPPL}$ (obtained from equation 1) for the main phases in samples S1-S3, respectively.

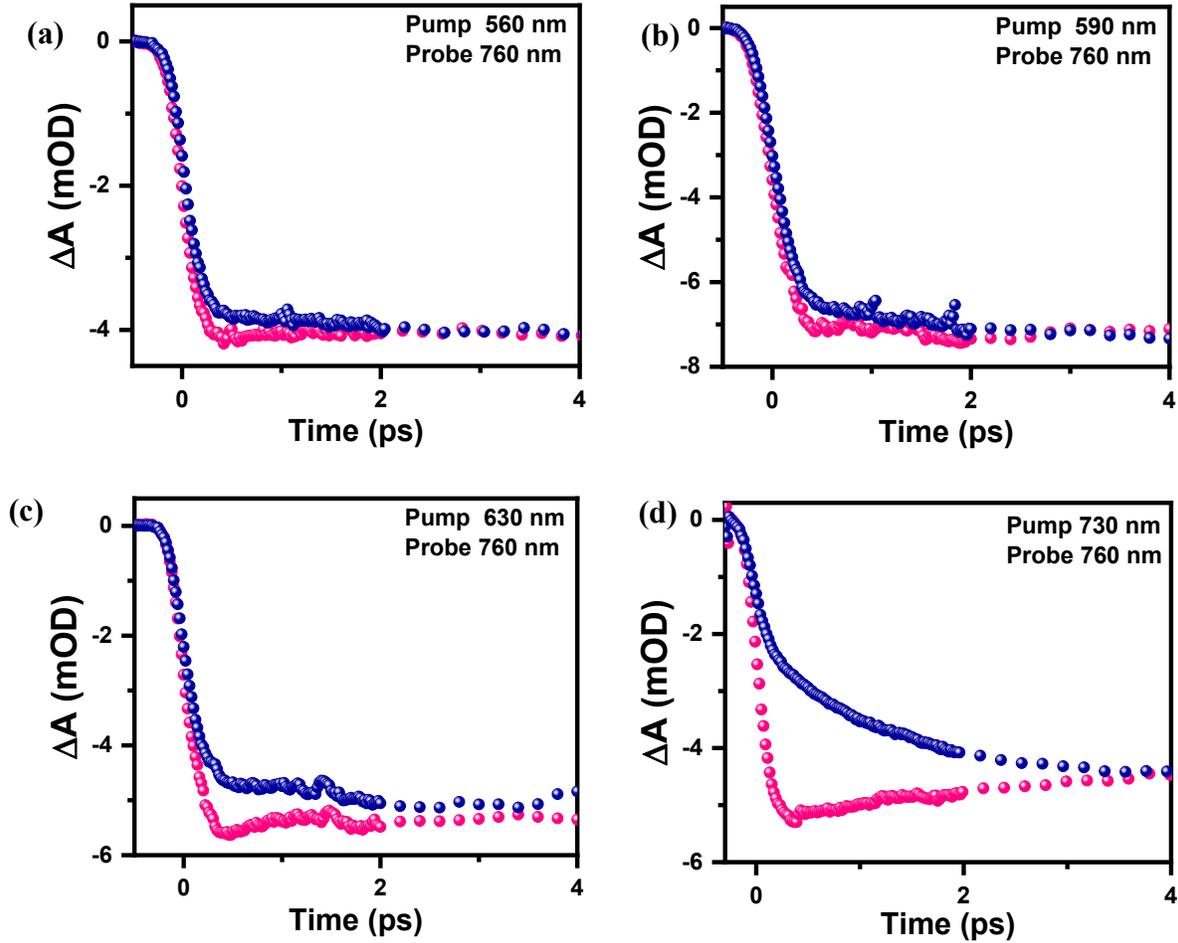

**Figure S3** TA kinetics measured at the bleach position of the bulk phase under band-edge excitation of **(a)** n=2, **(b)** n=3, **(c)** n=4, and **(d)** bulk phases. The pink and blue symbols correspond to same- and opposite-circular pump-probe polarization, respectively.

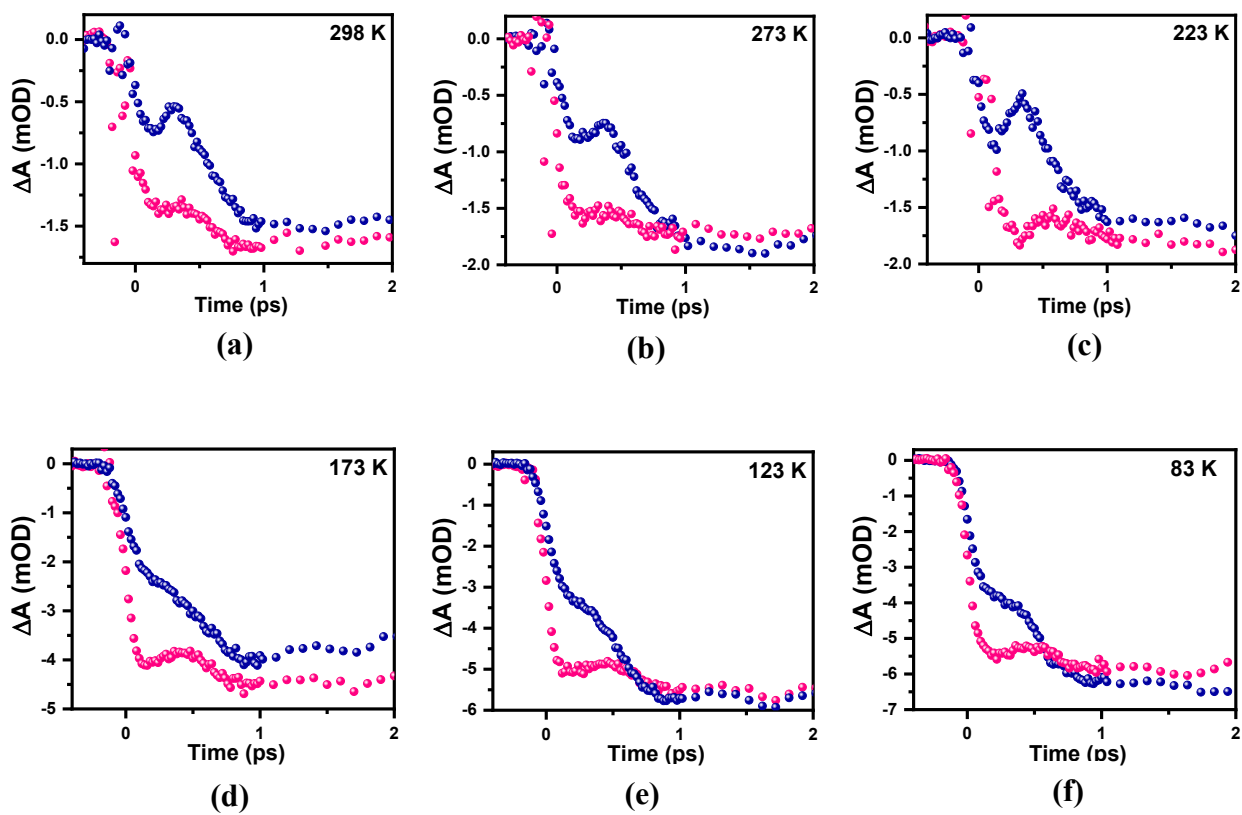

**Figure S4 (a-f)** Exciton bleach kinetics of n=1 phase under band-edge excitation (pump wavelength ~ 510 nm) at 298 K, 273 K, 223 K, 173 K, 123 K and 83 K, respectively. The pink and blue symbols correspond to same- and opposite-circular pump-probe polarization, respectively. The pump fluence was 2.5 μJ/cm$^2$.

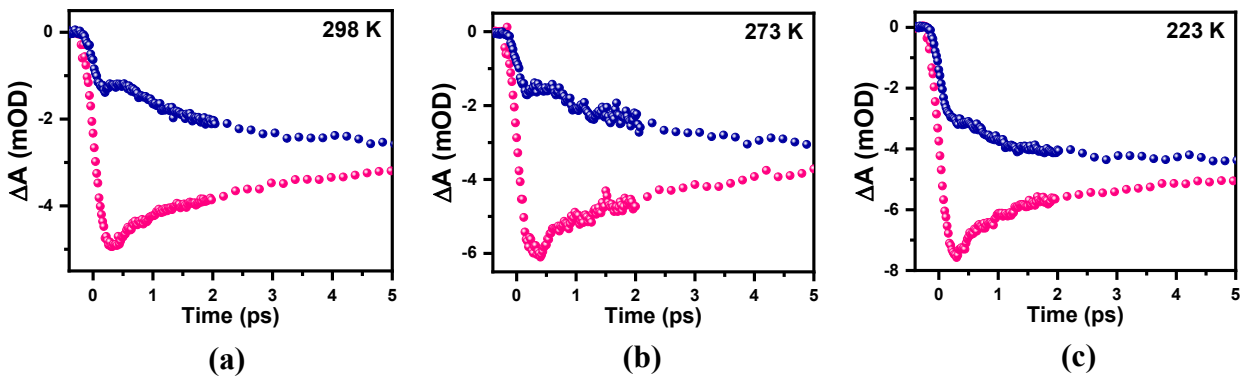

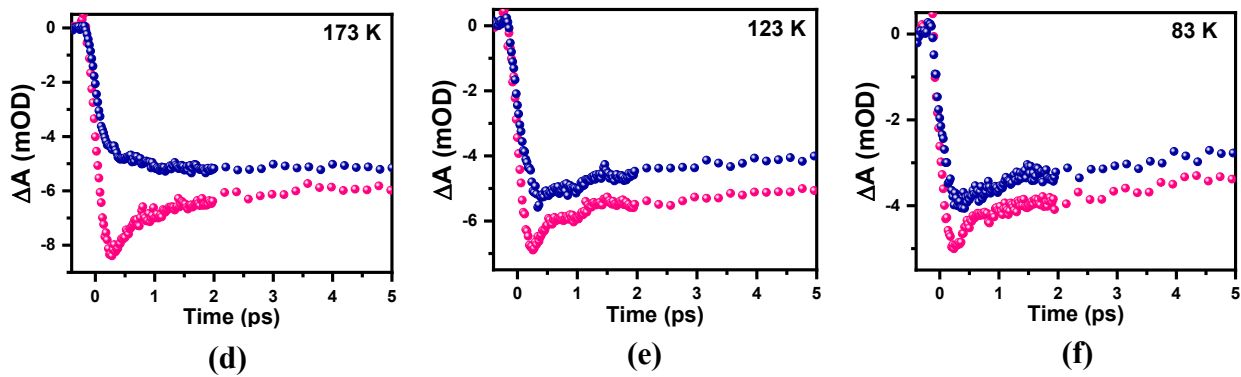

**Figure S5 (a-f)** Exciton bleach kinetics of n=2 phase under band-edge excitation (pump wavelength ~ 560 nm) at 298 K, 273 K, 223 K, 173 K, 123 K and 83 K, respectively. The pink and blue symbols correspond to same- and opposite-circular pump-probe polarization, respectively. The pump fluence was 2.3 μJ/cm$^2$.

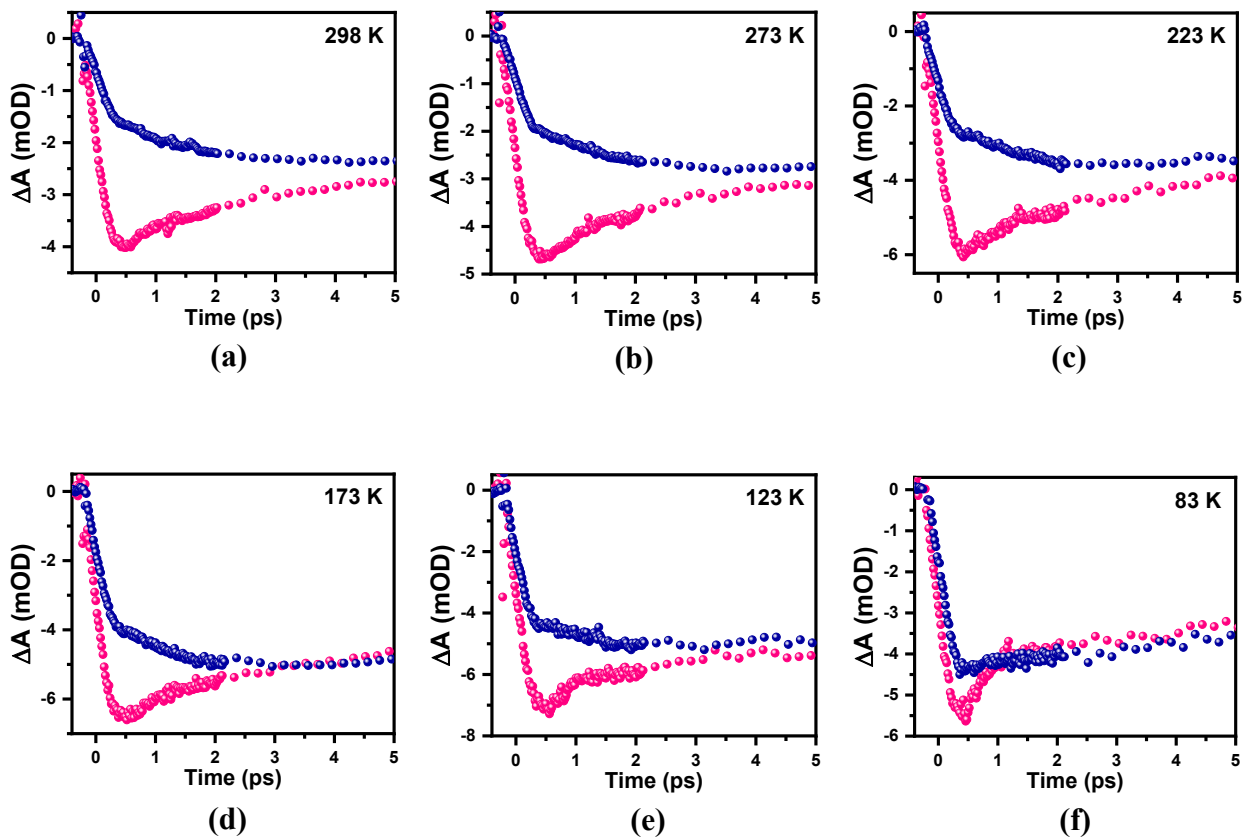

**Figure S6 (a-f)** Exciton bleach kinetics of n=3 phase under band-edge excitation (pump wavelength ~ 595 nm) at 298 K, 273 K, 223 K, 173 K, 123 K and 83 K, respectively. The pink and blue symbols correspond to same- and opposite-circular pump-probe polarization, respectively. The pump fluence was 10.6 µJ/cm$^2$.

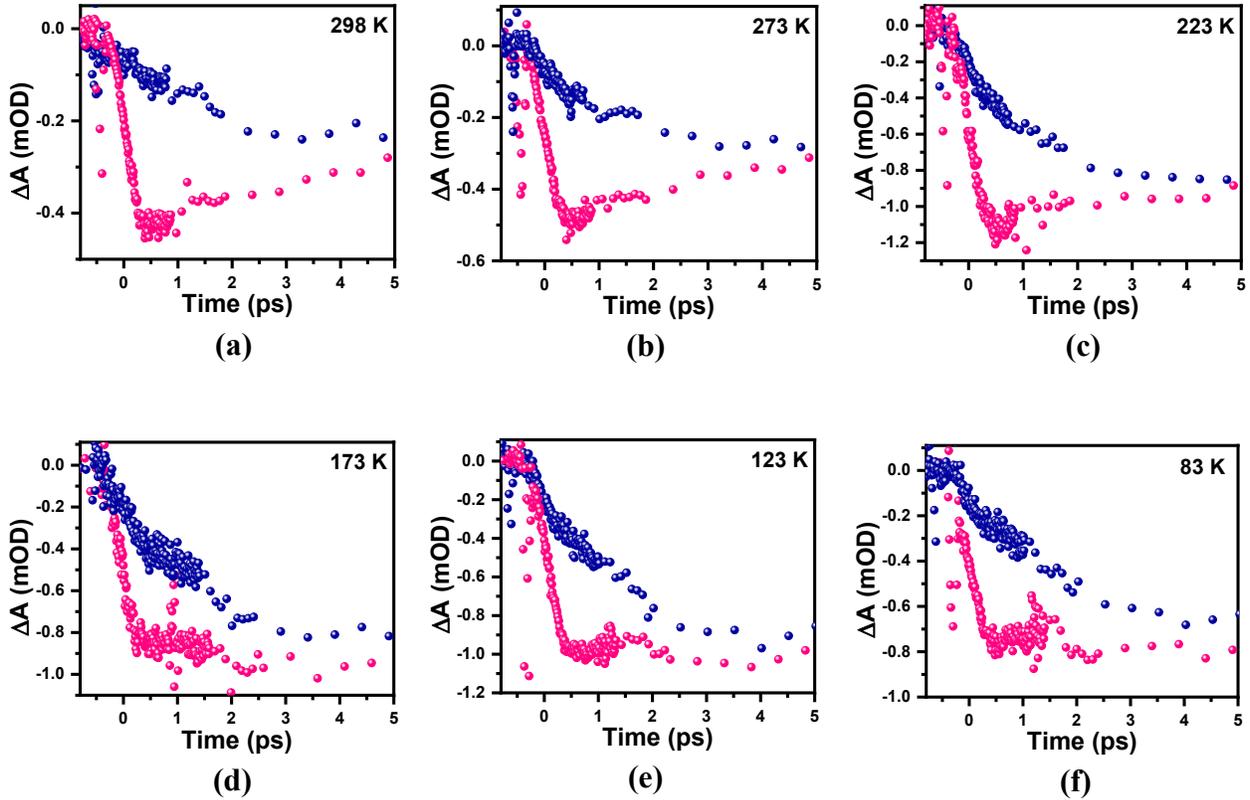

**Figure S7 (a-f)** Exciton bleach kinetics of n=4 phase under band-edge excitation (pump wavelength ~ 640 nm) at 298 K, 273 K, 223 K, 173 K, 123 K and 83 K respectively. The pink and blue symbols correspond to same- and opposite-circular pump-probe polarization, respectively. The pump fluence was 21.7 µJ/cm$^2$.

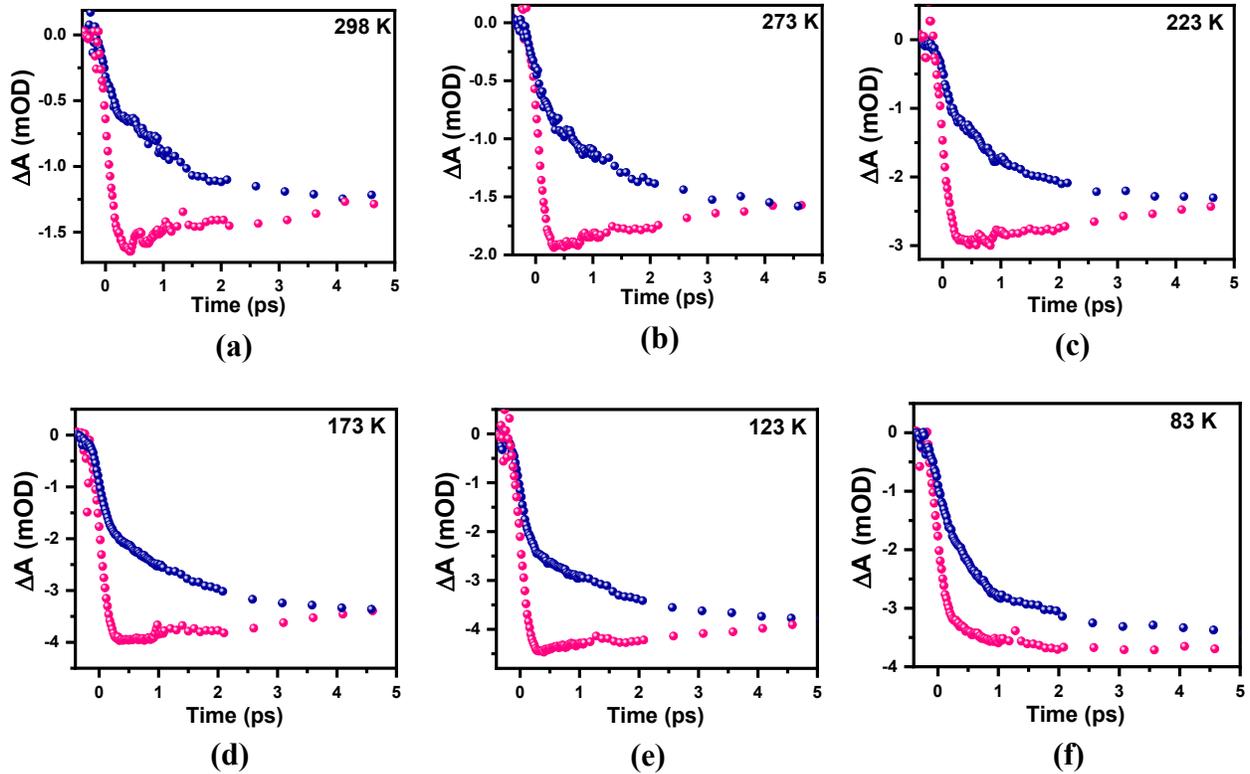

**Figure S8 (a-f)** Ground state bleach kinetics of bulk phase under band-edge excitation (pump wavelength ~ 730 nm) at 298 K, 273 K, 223 K, 173 K, 123 K and 83 K respectively. The pink and blue symbols correspond to same- and opposite-circular pump-probe polarization, respectively. The pump fluence was 25.8 µJ/cm$^2$.

| Temperature (K) | $\tau_S$ (ps) for n=1 | $\tau_S$ (ps) for n=2 | $\tau_S$ (ps) for n=3 | $\tau_S$ (ps) for n=4 | $\tau_S$ (ps) for bulk (3D) |
|---|---|---|---|---|---|
| 298 | 0.35 | 2.04 | 1.83 | 2.41 | 1.37 |
| 273 | 0.34 | 1.88 | 1.65 | 2.18 | 1.82 |
| 223 | 0.21 | 1.13 | 1.43 | 1.68 | 1.94 |
| 173 | 0.24 | 0.70 | 1.17 | - | 2.02 |
| 123 | 0.33 | 0.11 | 0.91 | 1.69 | 2.04 |
| 83 | 0.36 | 0.11 | 0.52 | 1.81 | 2.06 |

**Table S1** Spin lifetimes ($\tau_S$) for n=1-4 and bulk phases over the temperature range of 83-298 K, extracted by single-exponential fitting of the corresponding spin relaxation dynamics presented in **Figure 5**.

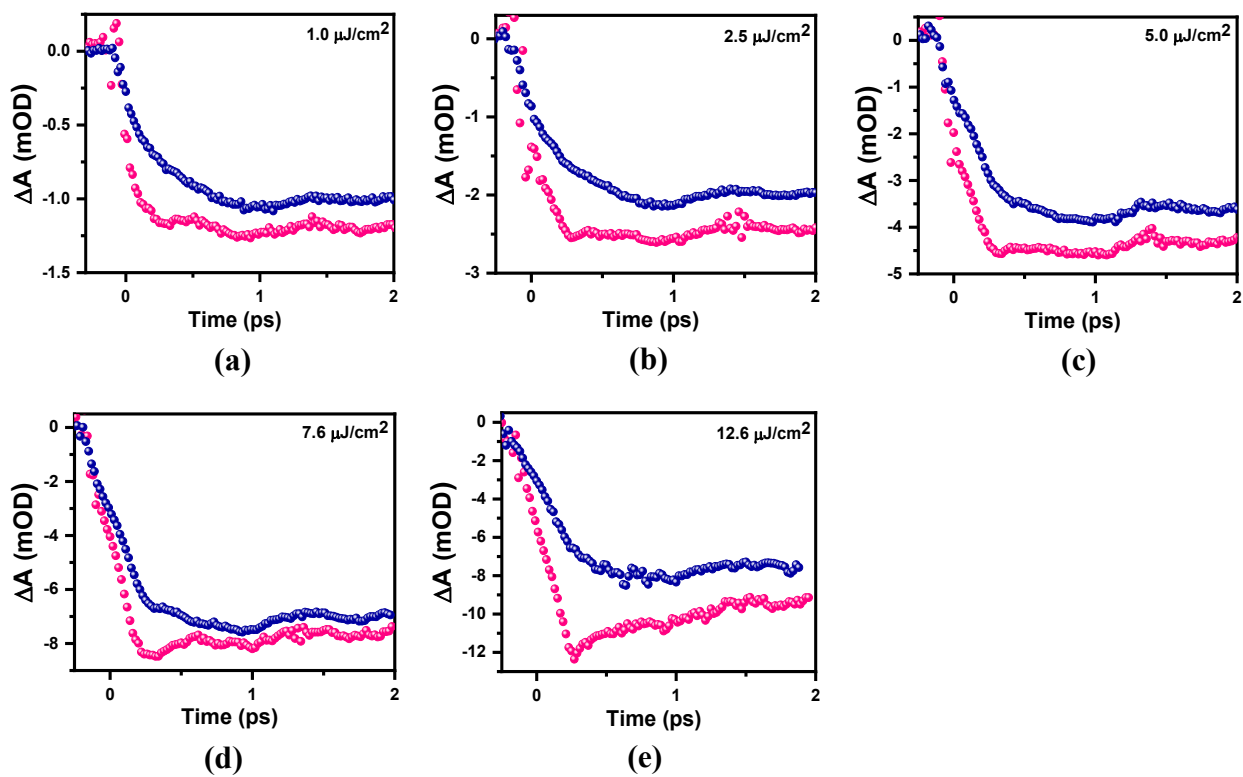

**Figure S9** Exciton bleach kinetics of n=1 phase under band-edge excitation (pump wavelength ~ 510 nm) at room temperature and different pump fluences. The pink and blue symbols correspond to same- and opposite-circular pump-probe polarization, respectively.

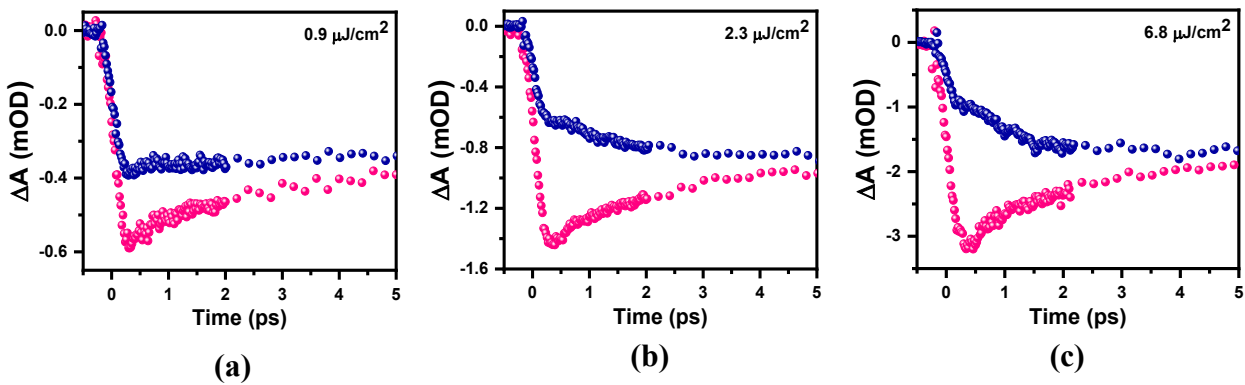

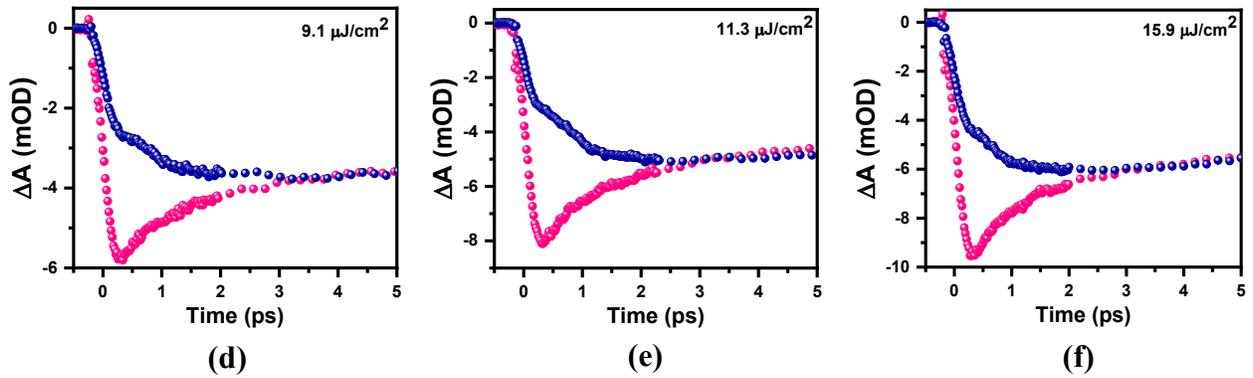

**Figure S10** Exciton bleach kinetics of n=2 phase under band-edge excitation (pump wavelength ~ 560 nm) at room temperature and different pump fluences. The pink and blue symbols correspond to same- and opposite-circular pump-probe polarization, respectively.

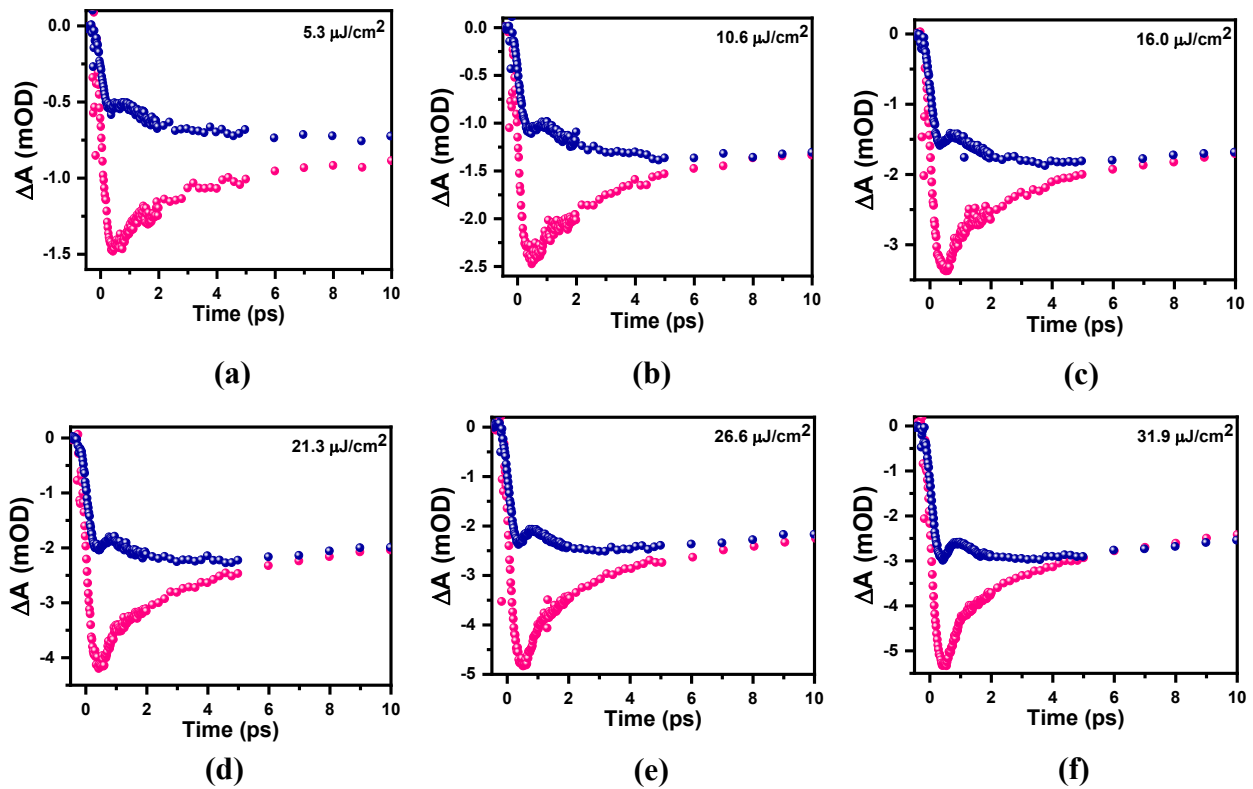

**Figure S11** Exciton bleach kinetics of n=3 phase under band-edge excitation (pump wavelength ~ 595 nm) at room temperature and different pump fluences. The pink and blue symbols correspond to same- and opposite-circular pump-probe polarization, respectively.

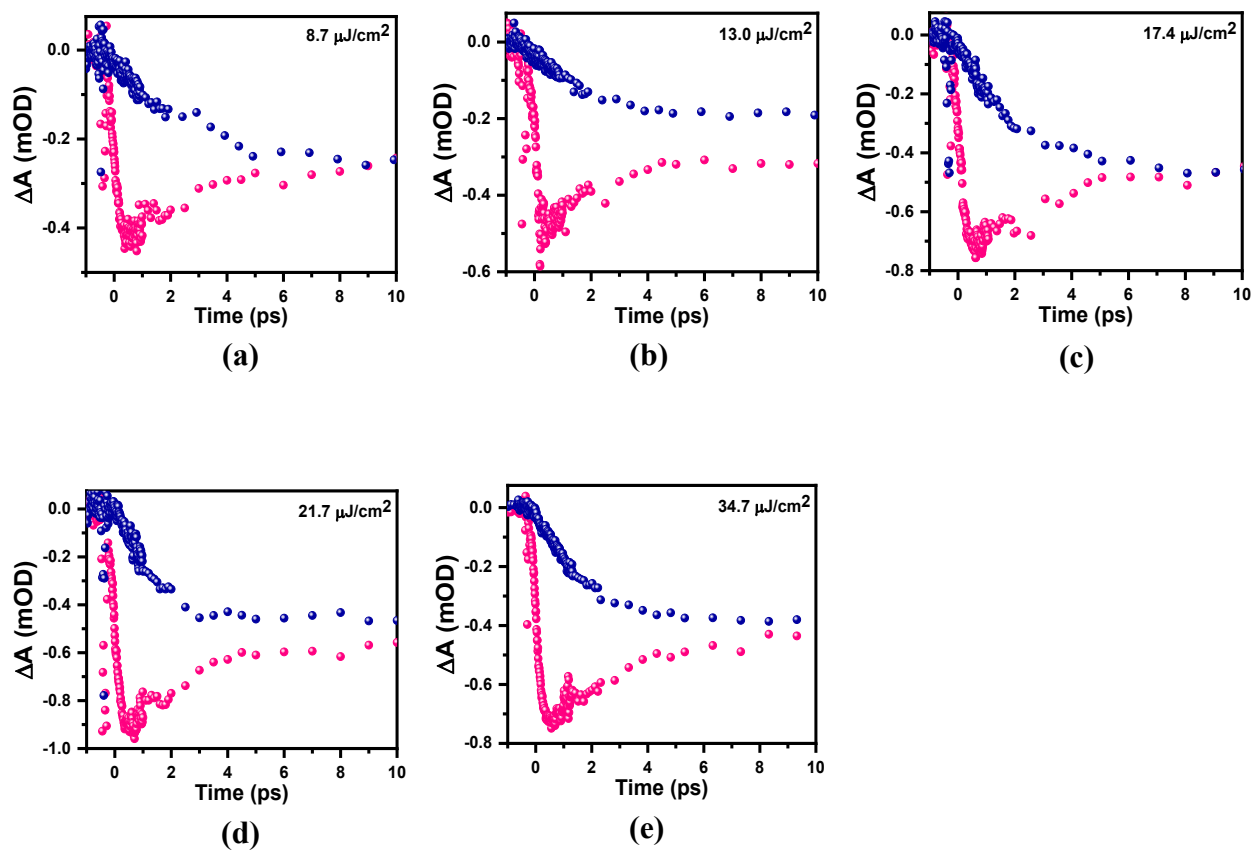

**Figure S12** Exciton bleach kinetics of n=4 phase under band-edge excitation (pump wavelength ~ 640 nm) at room temperature and different pump fluences. The pink and blue symbols correspond to same- and opposite-circular pump-probe polarization, respectively.

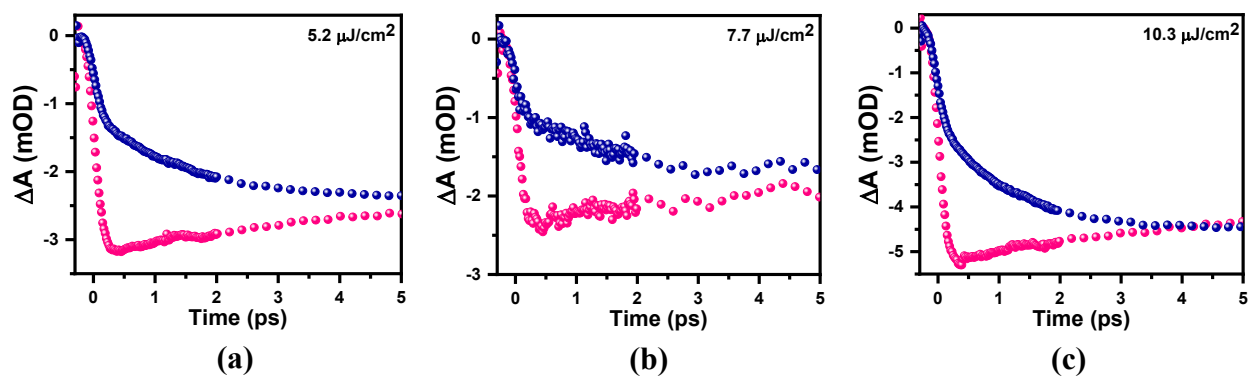

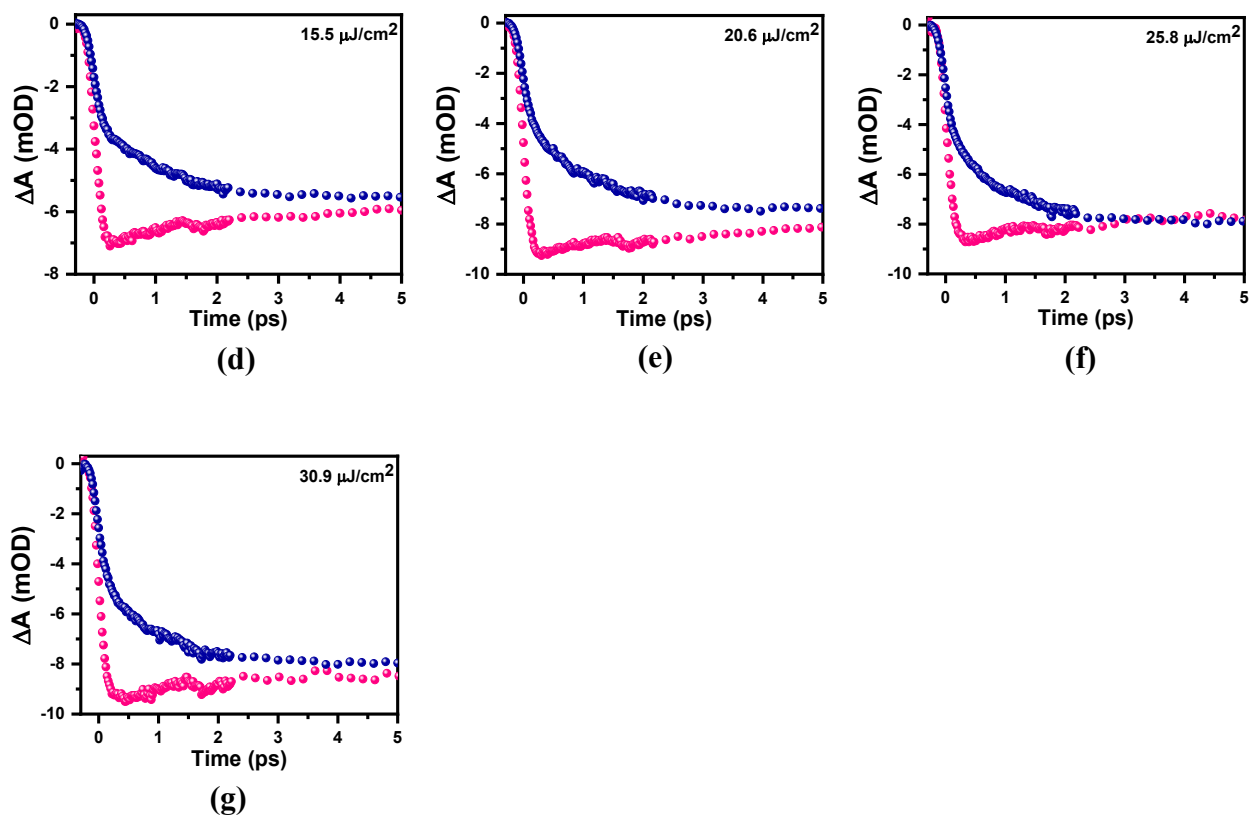

**Figure S13** Ground state bleach kinetics of bulk phase under band-edge excitation (pump wavelength ~ 730 nm) at room temperature and different pump fluences. The pink and blue symbols correspond to same- and opposite-circular pump-probe polarization, respectively.

| n=1 | | n=2 | | n=3 | |
|---|---|---|---|---|---|
| Pump Fluence ($\mu J/cm^2$) | $\tau_S$ (ps) | Pump Fluence ($\mu J/cm^2$) | $\tau_S$ (ps) | Pump Fluence ($\mu J/cm^2$) | $\tau_S$ (ps) |
| 1.0 | 0.20 | 0.9 | 2.77 | 5.3 | 2.39 |
| 2.5 | 0.22 | 2.3 | 1.75 | 10.6 | 2.20 |
| 5.0 | 0.21 | 6.8 | 1.36 | 16.0 | 1.69 |
| 7.6 | 0.22 | 9.1 | 0.99 | 21.3 | 1.71 |
| 12.6 | 0.28 | 11.3 | 0.84 | 26.6 | 1.40 |
|  |  | 15.9 | 0.74 | 31.9 | 1.36 |

| n=4 | | Bulk (3D) | |
|---|---|---|---|
| Pump Fluence ($\mu J/cm^2$) | $\tau_S$ (ps) | Pump Fluence ($\mu J/cm^2$) | $\tau_S$ (ps) |
| 8.7 | 3.54 | 5.2 | 1.96 |
| 13.0 | 2.47 | 7.7 | 1.87 |
| 17.4 | 2.80 | 10.3 | 1.61 |
| 21.7 | 2.23 | 15.5 | 1.29 |
| 34.7 | 2.41 | 20.6 | 1.30 |
|  |  | 25.8 | 1.09 |
|  |  | 30.9 | 1.13 |

**Table S2** Spin lifetimes ($\tau_S$) for n=1-4 and bulk phases at different pump fluences, extracted by single-exponential fitting of the corresponding spin relaxation dynamics presented in **Figure 6**.